\documentclass[universe,article,accept,moreauthors,pdftex]{Definitions/mdpi} 
\newcolumntype{C}[1]{>{\PreserveBackslash\centering}m{#1}}
\newcolumntype{R}[1]{>{\PreserveBackslash\raggedleft}m{#1}}
\newcolumntype{L}[1]{>{\PreserveBackslash\raggedright}m{#1}}
\graphicspath{{./figures/}}
\usepackage{relsize}
\usepackage{blindtext}
\usepackage{enumitem}
\usepackage{graphicx}
\usepackage{bm}
\usepackage{cancel}
\usepackage{epsfig}
\usepackage{dcolumn}
\usepackage{amsmath}
\usepackage{amssymb}
\usepackage{amsmath} 
\usepackage{hhline}
\usepackage{adjustbox}
\usepackage{enumerate}
\usepackage[utf8]{inputenc}
\usepackage{caption}
\usepackage{longtable}
\usepackage{rotating}
\usepackage{tablefootnote}
\usepackage{float}

\usepackage{nameref}

\def\aaps{{Astron.\@  \@Astroph.\ }}
\def\apss{{Astroph.\@ \& \@Space \@Science\ }}

\def\apj{{Astroph.\@ J.\ }}

\def \beq  {\begin{equation}}
\def \eeq  {\end{equation}}
\def \ber  {\begin{eqnarray}}
\def \eer  {\end{eqnarray}}

\newcommand{\newc}{\newcommand}

\newcommand{\be}{\begin{equation}}
\newcommand{\ee}{\end{equation}}
\newcommand{\ba}{\begin{eqnarray}}
\newcommand{\ea}{\end{eqnarray}}
\newcommand{\bea}{\begin{eqnarray*}}
\newcommand{\eea}{\end{eqnarray*}}
\newc{\D}{\partial}
\newc{\ie}{{i.e.,} }
\newc{\eg}{{e.g.,} }
\newc{\etc}{{etc.} }
\newc{\etal}{{et al.}}
\newc{\lcdm}{$\Lambda$CDM }
\newc{\lcdmnospace}{$\Lambda$CDM}
\newc{\wcdm}{$w$CDM }
\newc{\plcdm}{Planck18/$\Lambda$CDM }
\newc{\plcdmnospace}{Planck18/$\Lambda$CDM}
\newc{\omom}{$\Omega_{0m}$ }
\newc{\omomnospace}{$\Omega_{0m}$}

\newc{\ra}{\Rightarrow}
\newc{\baodv}{$\frac{D_V}{r_s}$ }
\newc{\baodvnospace}{$\frac{D_V}{r_s}$}
\newc{\baoda}{$\frac{D_A}{r_s}$ } 
\newc{\baodanospace}{$\frac{D_A}{r_s}$}
\newc{\baodh}{$\frac{D_H}{r_s}$ }
\newc{\baodhnospace}{$\frac{D_H}{r_s}$}

\firstpage{1} 
\pubvolume{1}
\issuenum{1}
\articlenumber{0}
\pubyear{2025}
\copyrightyear{2025}
\externaleditor{Firstname Lastname} 
\datereceived{ } 
\daterevised{ } 
\dateaccepted{ } 
\datepublished{ } 
\hreflink{https://doi.org/} 

\Title{Gravitational Wave Signatures Induced by Dark Fluid Accretion in Binary Systems}  
\TitleCitation{Gravitational Wave Signatures Induced by Dark  Fluid Accretion in Binary Systems}

\Author{Evangelos Achilleas Paraskevas *\orcidA{} and {Leandros Perivolaropoulos} \orcidB{}}
\AuthorNames{Evangelos Achilleas Paraskevas, Leandros Perivolaropoulos}
\AuthorCitation{Paraskevas, E.A.; Perivolaropoulos, L.}

\address[1]{Department of Physics, University of Ioannina, 45110 Ioannina, Greece; leandros@uoi.gr}

\corres{\hangafter=1 \hangindent=1.05em \hspace{-0.82em}Correspondence: e.paraskevas@uoi.gr}

\abstract{We investigate the impact of dark fluid accretion on gravitational waveforms emitted by a compact binary system consisting of a supermassive black hole and a stellar-mass black hole. Using a Lagrangian framework with 1~PN and 2.5~PN corrections, we analyze the effects of the spherically symmetric accretion of a fluid with steady-state flow, including those characterized by an equation of state parameter resembling dark energy, on the binary's dynamics.
 We validate our approach by comparing it with previous studies in the common region of validity and extend the analysis to include both local effects, such as dynamical friction, and global gravitational interactions with the stellar-mass black hole, focusing on their dependence on the fluid's properties.
Our analysis reveals that these interactions induce de-phasing in gravitational waveforms, with the phase shift influenced by the fluid's equation of state and energy density. We also extend the study to sudden cosmological singularities, finding that, although they can deform the binary's orbit from initially circular to elliptical, their effect on de-phasing is negligible for cosmologically relevant energy densities.
By incorporating both the local and global gravitational interactions of a fluid on a two-body system into the equations of motion, this preliminary study provides a framework for understanding the interplay between fluid dynamics and gravitational wave emissions in astrophysical systems.
 It further reinforces the potential for probing the properties of astrophysically relevant fluids through gravitational wave observations.}


\begin{document}

\section{Introduction}
\label{intro}
{The Advanced LIGO and Virgo interferometers detected approximately 90 events involving compact binary coalescences during their first three observing runs \citep{LIGOScientific:2018mvr,LIGOScientific:2020ibl,KAGRA:2021vkt,LIGOScientific:2021usb}. Furthermore, several proposed future missions aim to enhance gravitational wave detection, including the Laser Interferometer Space Antenna (LISA) \citep{LISA:2017pwj}, Taiji \citep{Hu:2017mde}, TianQin \citep{TianQin:2015yph}, the Einstein Telescope (ET) \citep{Punturo:2010zz}, and the DECi-hertz Interferometer Gravitational Wave Observatory (DECIGO) \citep{Seto:2001qf}. These missions will be crucial for exploring cosmology and testing modified gravity theories \citep{Poddar:2021yjd,Finke:2021znb,LISACosmologyWorkingGroup:2022wjo,Silva:2022srr,Banik:2023pbo,Loutrel:2022xok,Yang:2022fgp,Quartin:2023tpl,Chen:2024xkv}.}

{The observation of compact binary coalescences allows us to examine general relativity (GR) under extreme conditions. Gravitational waves generated by binary mergers, including those involving neutron stars, are vital for testing GR and directly probing the properties of matter under the extreme conditions present in the interiors of these stars~\mbox{\citep{LIGOScientific:2018cki,LIGOScientific:2018hze,LIGOScientific:2021qlt,Dietrich:2020efo,McLerran:2018hbz}.} Binary systems provide a compelling avenue for studying astrophysical environments, including the effects of dark matter on their orbital dynamics and evolution. For example, extreme mass ratio inspirals could plausibly indicate the presence of a spike of collisionless dark matter particles \citep{Eda:2013gg,Eda:2014kra,Speeney:2024mas}. Due to dynamical friction \citep{Chandrasekhar:1943ys}, this interaction is expected to cause a de-phasing in the gravitational waveform, an effect that future missions may be able to detect \citep{Li:2021pxf,Cole:2022ucw,Seoane:2021kkk}.}

{The dynamics of matter around black holes have been extensively explored, with a particular focus on stellar clusters and dark matter distributions. For example, collisional stellar cusps, as described by \citet{Bahcall:1976aa}, exhibit a steady-state density profile of $\rho(r) \sim r^{-7/4}$ consistent with observations of the Milky Way’s center. In cases of prolonged black hole growth, adiabatic compression alters the density profile to $\rho(r) \sim r^{-3/2}$, as shown by \citet{1972GReGr...3...63P}. \citet{Quinlan:1994ed} extended these analyses, demonstrating generalized power law behaviors following adiabatic growth. Further studies have extended these concepts to Schwarzschild and Kerr black holes, exploring their effects on dark matter density profiles \citep{Sadeghian:2013laa,Ferrer:2017xwm}.
While these models have been applied to dark matter spikes, dark matter profiles around black holes generally do not settle into a universal steady-state solution \citep{Ullio:2001fb,Merritt:2002vj}. However, \citet{Bertone:2005xz} and \citet{Zhao:2005zr} proposed that intermediate-mass black holes (IMBHs) are more likely to retain unperturbed dark matter spikes. In particular, dark matter mini-spikes are expected to persist around IMBHs. \citet{Bertone:2005xz} analyzed two scenarios for IMBH formation: (i) the formation of IMBHs in rare, overdense regions at high redshift ($z \sim 20$) as remnants of Population III stars, resulting in a $\rho \sim r^{-3/2}$ overdensity from adiabatic growth, and (ii) the formation of more massive IMBHs via direct collapse, which produces a steeper $\rho \sim r^{-7/3}$ profile.  }

{Interestingly, the presence of dark matter influences the evolution of binary black hole systems, primarily through the transfer of energy from the binary to the dark matter as a result of gravitational interactions between them.
Studies have explored the impact of dark matter on gravitational waveforms \mbox{\citep{Eda:2013gg,Eda:2014kra,Kavanagh:2020cfn,Coogan:2021uqv,Figueiredo:2023gas,Montalvo:2024iwq,Speeney:2024mas}}. \citet{Kavanagh:2020cfn} devised a  method to jointly evolve the binary in a distribution of dark matter, noting that the de-phasing induced by dark matter, when assuming a fixed dark matter density profile, tends to overestimate the de-phasing. For more information, refer to the recent review by \citet{Bertone:2024rxe}.}

{For a binary system composed of compact objects, at a distance from the center of mass of the system---in the order of a few times the orbital separation---gravity is already sufficiently weak. As a result, a previous study by \citet{Montalvo:2024iwq} presented a Lagrangian framework that integrates post-Newtonian (PN) corrections and includes dynamical friction from a dark matter spike (and possibly other environmental effects), in order to study gravitational wave emission from extreme mass ratio inspirals (EMRIs).
 Although for EMRIs, the PN expansion is insufficient, due to their highly relativistic nature, as the PN expansion converges slowly for velocities greater than approximately $v/c \sim 0.3$ (in strong-field regimes a particularly suitable method is the “self-force” approach) \citep{Maggiore:2018sht}, it was sufficient for our purposes, since we aimed to conduct a preliminary study on the effects of various fluids. Therefore, we focus on regions sufficiently distant from the strong-field regime over relatively short time intervals, while investigating the relative de-phasing by incorporating various physical effects into the equations of motion.

  The framework accounts for effects such as dynamical friction, accretion, and other orbital or environmental factors in compact binary systems. 
As long as these effects are expressed as dissipative power or force, they can be incorporated as generalized forces \citep{Montalvo:2024iwq}. The Euler--Lagrange equations then yield the modified orbits, from which the gravitational waveforms emitted by the system are derived. }
{In the context of the approximations made in the previous study \cite{Montalvo:2024iwq}, a static dark matter distribution around the supermassive black hole was assumed. The supermassive black hole was modeled as a Schwarzschild black hole (BH) with a mass that grew adiabatically, forming a surrounding dark matter (DM) spike~\citep{Speeney:2022ryg} from an initial Navarro--Frenk--White (NFW) profile \citep{Navarro:1996gj}. The study also considered the local effect of dynamical friction \citep{Chandrasekhar:1943ys,Petrich:1988zz} from the dark matter spike on a stellar-mass black hole orbiting the supermassive one. }

{We have adopted a framework similar to \citet{Montalvo:2024iwq} to study EMRI binary systems, modeling the supermassive black hole as a Schwarzschild black hole surrounded by a dark fluid, which is at rest at infinity. The dark fluid accretes onto the black hole as described by \citet{Babichev:2004yx}. \citet{Montalvo:2024iwq} studied a static dark matter spike, neglecting steady-state velocities, despite their expected presence in Schwarzschild spacetime and their relevance in fluid dynamics \citep{Babichev:2004yx}.   The simplest case of spherically
symmetric stationary accretion involves the stationary, spherically symmetric solution first discussed by \mbox{\citet{Bondi:1952ni}}, where an infinitely large homogeneous gas cloud steadily accretes onto a central gravitational object, formulated within Newtonian gravity. Later, in the framework of general relativity (GR), \citet{Michel:1972oeq} investigated the steady-state spherically symmetric flow of test fluids (polytropic gas) onto a Schwarzschild black hole.  Since then, spherical accretion has been extensively studied for various  (dark) fluids static black holes in GR and modified gravities \cite{1980ApJ...235.1038M,Babichev:2004yx,Babichev:2013vji,Bahamonde:2015uwa,Chaverra:2015bya,Chaverra:2015oqb,Jawad:2016fop,Yang:2020bpj,10.1093/mnras/stab1127,Gupta:2024gun}. It should be noted that for any given positive particle density at infinity, along with vanishing velocity at infinity, there exists a unique steady-state radial accretion flow of fluid into a non-rotating black hole, which remains regular at the horizon, when considering steady radial accretion \citet{Chaverra:2015bya}. A key feature of spherical accretion onto black holes is transonic accretion and the existence of the critical point (sonic point), where the accretion flow transitions from subsonic to supersonic. Typically, the self-gravity of the accreting fluid is neglected, although some authors have considered it \citep{Malec:1999dd}.}

 By modeling the supermassive black hole as a Schwarzschild black hole, we assume spherical accretion of the dark fluid and we investigate the resulting density profiles and velocities based on its equation of state (EoS). This allows us to explore its effects on stellar compact objects, accounting for gravitational influences and identifying potential observables for future missions. We examine steady-state density distributions of fluids, modeled under the test-fluid approximation (the fluid moves under the black hole's gravitational field while its own gravitational field is ignored), with an equation of state (EoS) $p = \alpha (\rho - \rho_0)$ \citep{Babichev:2004qp}, considering cosmologically relevant energy densities of the order $\rho_{\rm crit} = 3 H_0^2/8 \pi G$ \citep{Planck:2018vyg}. 
 This study incorporates into the equations of motion of the binary the local dissipative effects of the fluid on the stellar compact object along with its global gravitational influence, which is included as an additional correction to the already 1~PN-corrected gravitational field of the supermassive black hole. Based on the derived density and velocity profiles, the global gravitational interaction of the fluid with the stellar compact object is induced by a spherical shell extending from the Schwarzschild radius to the orbital radius of the stellar compact object.
Since this study focuses on the binary system over relatively short time intervals, when its components are widely separated, the analysis is simplified by neglecting spin--orbit and spin--spin interactions (see \citet{Hartl:2004xr}, \citet{Barausse:2009xi}).

{The questions addressed in the present analysis are the following:}
\begin{itemize}
  
\item  How do dark fluids influence binary trajectories through dissipation and modifications to gravitational interaction at the post-Newtonian level, and how do these changes affect the resulting gravitational waveforms?
\item What is the magnitude of phase modifications due to dark fluids compared to those
anticipated from post-Newtonian effects in the context of general relativity?
    \item How does the phase of gravitational waves depend on the equation of state of dark fluids?
    \item  Can gravitational waves be used to probe the properties of dark fluids around black holes or neutron stars?
    \end{itemize}

{The structure of this paper is as follows: First, we validate the results presented in~\cite{Montalvo:2024iwq}, by incorporating both the 1~PN and 2.5~PN terms. Next, we adopt the methodology of \citet{Babichev:2013vji} to analyze steady-state flows in a static Schwarzschild background for dark fluids, including exotic equations of states such as those resembling dark energy. We then investigate the impact of these fluids on the stellar compact object through both local and global gravitational interactions: locally, via dynamical friction, and globally, through the gravitational influence of a spherical shell surrounding the supermassive black hole. Our focus is on how these forces are affected by the fluid's equation of state. Subsequently, we examine the correlation between variations in the fluid's equation of state and the resulting de-phasing of gravitational waves induced by a binary within such a fluid, compared to an identical binary in the absence of a fluid, to determine whether these effects could lead to potential observable signatures.
Finally, the framework presented enables us to extend our analysis, aiding in the interpretation of the effects of a sudden cosmological singularity \citep{Barrow:2004xh}, particularly concerning the divergence of pressure at the moment of the singularity's occurrence.}

\section{Equations of Motion with Relativistic Corrections Using the Lagrangian Method}

Consider an isolated binary system consisting of two compact objects, such as black holes or neutron stars. Let \(m_1\) represent the mass of one object, while \(m_2\) denotes the mass of the other compact object.
Furthermore, the total mass of the two-body system is denoted by \(m = m_1 + m_2\), and its reduced mass by \(\mu = \frac{m_1m_2}{m}\). Also, we denote as 
 \(\mathbf{r} = \mathbf{r}_1 - \mathbf{r}_2\)  the relative position of the two objects, with $\mathbf{r}_1$ and $\mathbf{r}_2$ referring to the central and the orbiting objects, respectively. 
  In spherical coordinates, the position vector $\mathbf{r}$ is expressed as
\[
\mathbf{r} = \left( x, y, z \right) = \left( r \cos(\phi) \sin(\theta), r \sin(\phi) \sin(\theta), r \cos(\theta) \right).
\]

If the motion is restricted to a two-dimensional plane, with \( \theta = \frac{\pi}{2} \), the position simplifies to
\[
\mathbf{r} = \left( x, y \right) = \left( r \cos(\phi), r \sin(\phi) \right),
\]
and the velocity becomes
\[
\mathbf{v} \equiv \dot{\mathbf{r}} = \dot{r} \mathbf{\hat{r}} + r \dot{\phi} \boldsymbol{\hat{\phi}}.
\]

The Lagrangian for the  Newtonian part is given by
\begin{equation}
L = \frac{1}{2} \mu v^2 + \frac{G\mu m}{r}. 
\end{equation}

In the context of the Two-Body Problem, using the Post-Newtonian approximation (applied to a system of slowly moving particles bound by gravitational forces), the acceleration in the center of mass (CM) frame for two particles can be written as (refer to \cite{Pati:2000vt,Maggiore:2007ulw,Will:2011nz,Straumann:2013spu,Blanchet:2013haa} for further details)
\begin{equation}
\frac{d\mathbf{v}}{dt} = \frac{Gm}{r^2} \left(-\mathbf{\hat{r}}+ \frac{1}{c^2}\mathbf{A}_{1\text{PN}}+ \frac{1}{c^4}\mathbf{A}_{2\text{PN}} + \frac{1}{c^5}\mathbf{A}_{2.5\text{PN}} +  \ldots \right),
\end{equation}
The term at the first post-Newtonian (1~PN) order represents the first general relativistic correction and corresponds to a conservative force described by
 \citep{Pati:2002ux}:
\begin{equation}
\mathbf{A}_{1\text{PN}} = \left[ (4 + 2\eta) \frac{Gm}{r} - (1 + 3\eta)v^2 + \frac{3}{2} \eta \dot{r}^2 \right] \mathbf{\hat{r}} + (4 - 2\eta)\dot{r} v \mathbf{\hat{v}},
\end{equation}
where $\eta = \frac{\mu}{m}$. This term accounts for the precession of orbits.

Additionally, the 2.5~PN-order, $A_{\rm 2.5~PN}$, dictates the orbital decay due to the emission of gravitational waves (back-reaction of GWs).  This term corresponds to a dissipative force and is described in \citep{Pati:2002ux}:
\begin{equation}
\mathbf{A}_{2.5\text{PN}} = \frac{8}{15} \eta \frac{Gm}{r} \left[ \left(9v^2 + \frac{17Gm}{r} \right) \dot{r} \mathbf{\hat{r}} - \left(3v^3 + \frac{9Gm}{r}v \right) \mathbf{\hat{v}} \right].
\end{equation}

Here, instead of following the standard procedure of adding the conservative 1~PN term to obtain the Lagrangian equations of motion, i.e., using the effective one body (EOB) action (see \cite{Buonanno:2000ef,Blanchet:2013haa,Maggiore:2018sht}), we simplify the process by accounting for the 1~PN correction through the methodology outlined in \citet{Montalvo:2024iwq}.
 This involves directly implementing the 1~PN and 2.5~PN forces, corresponding to post-Newtonian corrections, as generalized forces into the Euler--Lagrange equations.
The corresponding force can be expressed as follows
 \citep{Montalvo:2024iwq}:
\begin{equation}
F = F_r \mathbf{\hat{r}} + F_v\mathbf{\hat{v}}\,,
\end{equation}
where $F_r$ and $F_v$ are the radial and velocity-tangential components of the force, respectively.
The generalized forces $Q_r$ and $Q_\phi$ corresponding to such a force, expressed in terms of $\mathbf{\hat{r}}$ and $\mathbf{\hat{v}}$, are given by
\begin{equation}
Q_{q_j} = F \cdot \frac{\partial \mathbf{r}}{\partial q_j} = F_r \left( x \frac{\partial x}{\partial q_j} + y \frac{\partial y}{\partial q_j} \right) + F_v \left( \dot{x} \frac{\partial \dot{x}}{\partial \dot{q}_j} + \dot{y} \frac{\partial \dot{y}}{\partial \dot{q}_j} \right),
\end{equation}
where we use the relation $\frac{\partial \mathbf{r}}{\partial q_j} = \frac{\partial \dot{\mathbf{r}}}{\partial \dot{q}_j}$. In this expression, $q_1 = r$ and $q_2 = \phi$ represent the radial and angular coordinates. Using $v = \sqrt{\dot{r}^2 + r^2 \dot{\phi}^2}$, the radial and angular parts of the generalized force $Q_{q_j}$ are as follows:
\begin{equation}\label{qr}
    Q_r = F_r+ F_v \dot{r} \left( \dot{r}^2 + r^2 \dot{\phi}^2 \right)^{-1/2},
\end{equation}
\begin{equation}\label{qphi}
    Q_{\phi} = F_v r^2 \dot{\phi} \left( \dot{r}^2 + r^2 \dot{\phi}^2 \right)^{-1/2}.
\end{equation}

The Lagrangian equations of motion, including the post-Newtonian corrections as generalized forces, are written as follows:
\begin{equation}
    \frac{d}{dt} \left( \frac{\partial L}{\partial \dot{r}} \right) - \frac{\partial L}{\partial r} = Q^{\rm 1~PN}_{r} +Q^{\rm 2.5~PN}_{r}  ,
\end{equation}
\begin{equation}
    \frac{d}{dt} \left( \frac{\partial L}{\partial \dot{\phi}} \right) - \frac{\partial L}{\partial \phi} = Q^{\rm 1~PN}_{\phi}+Q^{\rm 2.5~PN}_{\phi}  .
\end{equation}

After dividing by $\mu$, we re-scale the differential equations by defining the dimensionless variables:
\begin{equation}\label{rescaledq}
    \bar{r} \equiv \frac{r}{c t_{\text{yr}} e^2}, \quad \bar{t} \equiv \frac{t}{ t_{\text{yr}} e},
\end{equation}
where $t_{\rm yr} = 1 \text{ yr}$ denotes one year, and $e$ denotes a dimensionless parameter\endnote{{Interestingly, the proper time for an object to undergo radial free fall from rest at the event horizon of a Schwarzschild black hole with mass \( m \) to the curvature singularity is \( \pi e^4 \) years.}}, defined as
\begin{equation}
    e \equiv \left(\frac{G m}{c^3 t_{\text{yr}}}\right)^{\frac{1}{4}}.
\end{equation}
The equations of motion are expressed perturbatively in terms of $e$ \citep{Montalvo:2024iwq}:
\begin{align}\label{eqofmotionr}
    \bar{r}'' - \bar{r}\phi'^2 + \frac{1}{\bar{r}^2} &= e^2\left[\frac{2}{\bar{r}^3}(2+\eta) + \frac{(6-7\eta)\bar{r}'^2}{2\bar{r}^2} - (1+3\eta)\phi'^2\right] \nonumber \\
    &\quad + e^5\left[16\eta\frac{\bar{r}'}{\bar{r}^3}\left(\bar{r}'^2+\bar{r}^2\phi'^2\right)+\frac{64\eta \bar{r}'}{15\bar{r}^4}\right],
\end{align}
\begin{equation}\label{eqofmotionphi}
\begin{aligned}
    \bar{r}^2\phi'' + 2\bar{r}\bar{r}'\phi' &= e^2\left[2(2-\eta)\bar{r}'\phi'\right]-e^5\left[\frac{8\eta\phi'}{5\bar{r}}(\bar{r}'^2+\bar{r}^2\phi'^2)+\frac{24\eta\phi'}{5\bar{r}^2}\right]\\
    \end{aligned}
\end{equation}
where the prime notation $' \equiv d/d\bar{t}$ is used for simplicity.

Here, the $e^2$-term corresponds to the 1~PN order, and the $e^5$-term corresponds to the 2.5~PN order, as the differential equations are expressed perturbatively in terms of $e$. Given that these terms are sufficient to analyze the orbit and evolution of binary pulsars \citep{Will:2011nz}, and since we focus on EMRI binaries \citep{Montalvo:2024iwq} at relatively large distances over small time intervals without considering their merger, we will neglect the 2PN terms to significantly simplify the differential equations we need to solve. This is justified, as both the 1~PN and 2PN terms are conservative and do not contribute to orbital decay. However, the first non-conservative effect, associated with gravitational radiation reaction, occurs at the 2.5~PN level. These terms remain important as they represent the leading radiation reaction effect and dictate the orbital decay, introducing a new feature to the dynamics.

We initialize the orbit at $r_{\text{init}}$ with $\dot{r}_{\text{init}} = 0$, assuming that the binary black holes start sufficiently far apart.
The inclusion of the 1~PN correction introduces a small shift in the potential's minimum relative to the Newtonian case (if the orbit were initiated from the Newtonian minimum, the radial component would exhibit oscillations around the slightly displaced new minimum).  To ensure the orbit starts at this new minimum, the re-scaled initial angular velocity is given by
\begin{equation}\label{omegainit}
\phi'_{\text{init}} = \frac{\sqrt{  2 (2 +\eta) e^2 - \bar{r}_{\rm init}}}{\bar{r}_{\rm init}^{3/2} \sqrt{ e^2 (1+3\eta)- \bar{r}_{\rm init}}}
\end{equation}
In the special case where the 1~PN term is neglected, the initial angular velocity simplifies to $\phi'_{\text{init}} = 1 / \bar{r}^{3/2}_{\text{init}}$, which corresponds to the Newtonian value.
Meanwhile, gravitational wave dissipation effects from the $e^5$ term gradually accumulate, leading to orbital decay over time. Physically, this results in a quasi-circular inspiral phase, governed by the 2.5~PN order, which accounts for the back-reaction on the motion, due to gravitational wave \mbox{(GW) emission. }

\begin{figure}[H]%

\includegraphics[width =0.9 \columnwidth]{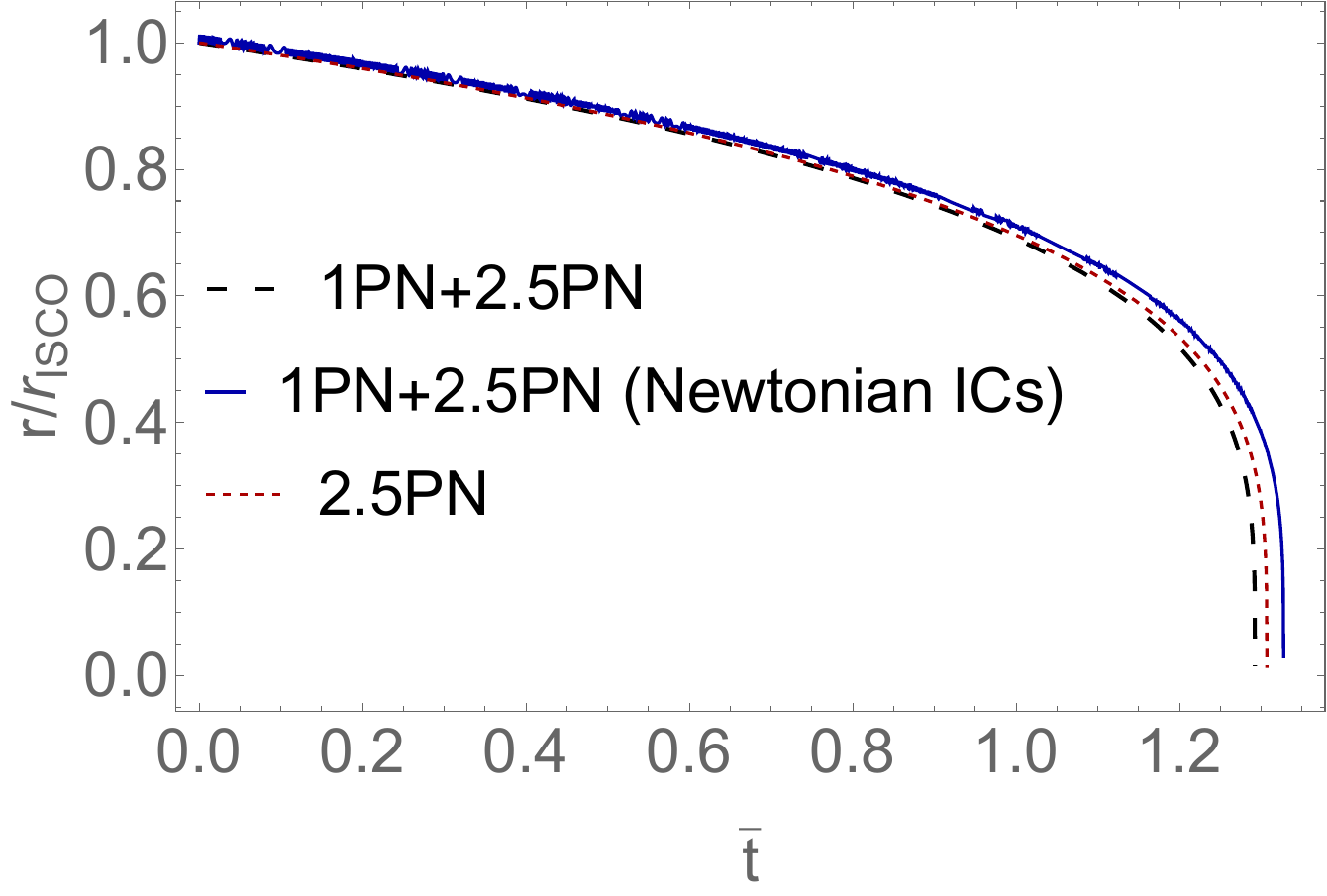}
\includegraphics[width =0.9 \columnwidth]{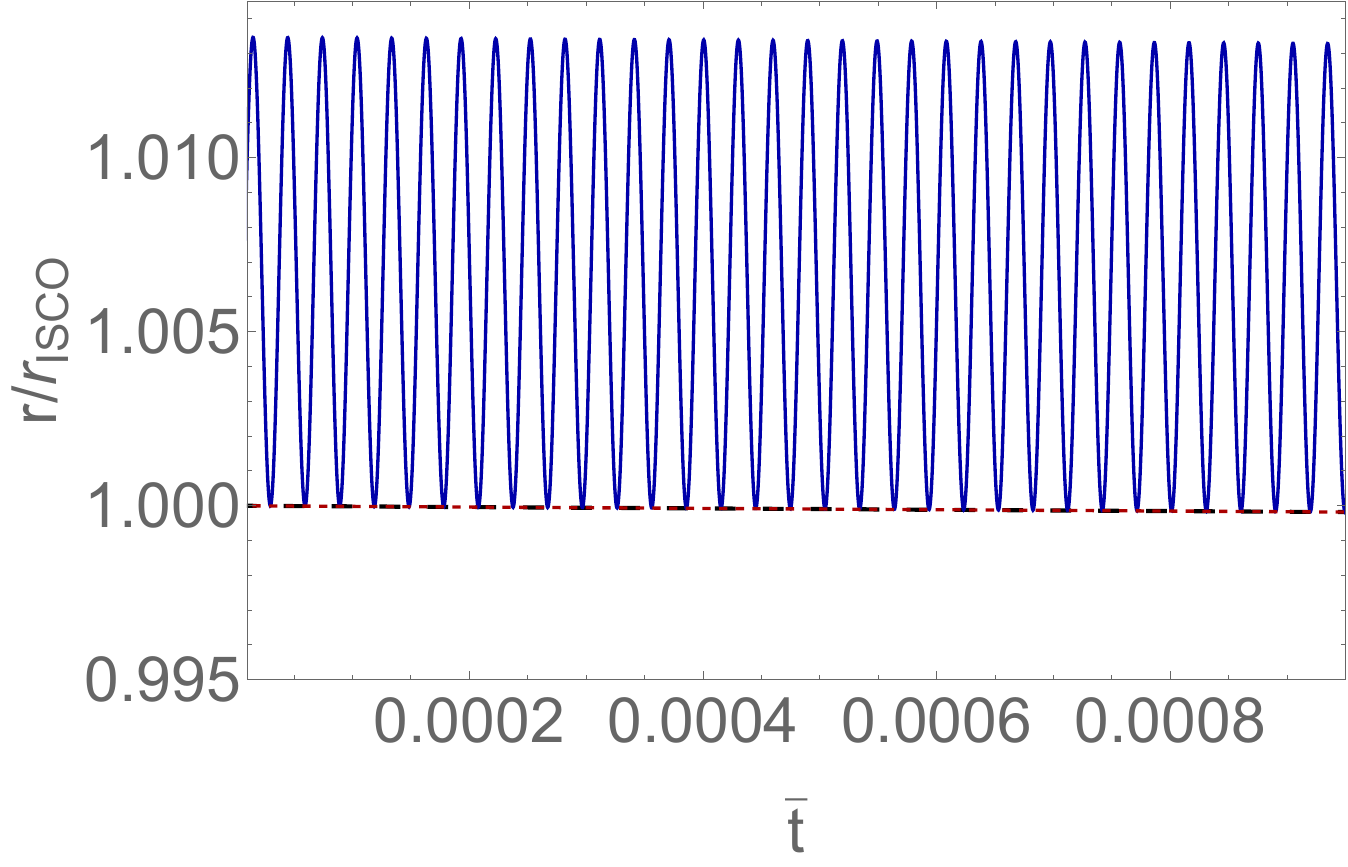}
\caption{The plot shows $r/r_{\rm ISCO}$ as a function of re-scaled time (see Equation~(\ref{rescaledq})), with the lower panel providing a magnified view of the upper panel. For this, we study a binary system with component masses $m_1 = m_2 = 1.4 M_{\odot}$. The system's reduced mass $\mu$ starts at an initial separation of $r_{\rm init} = 70 r_{\rm ISCO}$ in the center of mass (CM) frame.
 The equations of motion (Equations~(\ref{eqofmotionr})--(\ref{eqofmotionphi})) are solved. We compare the evolution of  $r/r_{\rm ISCO}$ with the case where the 1~PN correction is neglected, including only the 2.5~PN correction. The initial conditions are crucial; when the binary system begins at the minimum of the Newtonian potential (referred to as Newtonian initial conditions or Newtonian ICs), the inclusion of the 1~PN correction results in a longer merger time compared to the case where the 1~PN term is omitted.
However, when the initial conditions are set using Equation~(\ref{omegainit}), corresponding to the minimum of the effective potential, the inclusion of the 1~PN correction results in a slightly shorter merger time.}
\label{fig:0}
\end{figure}

\textls[-15]{For example, in Figure~\ref{fig:0} we present a plot of $r/r_{\rm ISCO}$,  where $r_{\rm ISCO}$ denotes the Schwarzschild value of the innermost stable circular orbit, as a function of re-scaled time (see Equation~(\ref{rescaledq})). For this, we consider a binary system with component masses \mbox{\( m_1 = m_2 = 1.4\,M_{\odot} \),} with the reduced mass \( \mu \) starting at an initial separation of \mbox{\( r_{\rm init} = 70\,r_{\rm ISCO} \)} in the center of mass (CM) frame. The equations of motion (\mbox{Equations~(\ref{eqofmotionr})--(\ref{eqofmotionphi})})} are solved.  We compare the radial evolution up to the merger for two scenarios: one that includes both the 1~PN and 2.5~PN corrections, and another that incorporates only the 2.5~PN correction, neglecting the 1~PN term.  The initial conditions are crucial \citep{Zwick:2019yjl};  when the binary systems start with Newtonian initial conditions, the inclusion of the 1~PN correction extends the merger time to $9.46$ h, compared to $9.32$ h when the 1~PN term is excluded.
However, when the initial conditions are derived from Equation~(\ref{omegainit}), corresponding to the minimum of the effective potential, the inclusion of the 1~PN correction slightly reduces the merger time, justifying the omission of the 2PN terms, as their effect would be even smaller. The merger time is reduced to approximately $9.21$ h.

Given Equation~(\ref{omegainit}), which determines the value of $\dot{\phi}_{\rm init}$ that minimizes the effective potential when the 1~PN correction is included, $\dot{\phi}_{\rm init}$ attains a lower value compared to the case where the 1~PN correction is not included.  Starting from Newtonian initial conditions produces an elliptical orbit whose semi-minor axis coincides with the radius of the circular orbit, leading to oscillations around the true minimum of the effective potential. The radial oscillations gradually diminish over time, and the elliptical orbit gradually becomes more circular. Including the 1~PN correction to the gravitational field reduces the $\dot{\phi}_{\rm init}$ required to sustain a circular orbit. Starting with the Newtonian $\dot{\phi}_{\rm init}$ results in excess energy that must be radiated away over time, leading to a slowing of the orbital decay.

For an isolated system (i.e., with vanishing acceleration of the center of mass) and the origin of the coordinate system at the center of mass, the center of mass motion does not contribute to gravitational wave emission.
Consequently,  in the center-of-mass frame, the mass density can then be expressed as
\[
\rho(t, \mathbf{x}') = \mu \delta^{(3)} \big( \mathbf{x}' - \mathbf{x}(t) \big)=\mu \delta(x' - x(t)) \delta(y' - y(t)) \delta(z'),
\]
where \(\delta^{(3)}\) is the three-dimensional Dirac delta function centered at \(\mathbf{x}(t)\).
Solving the equations of motion enables the explicit calculation of waveforms as functions of time. Once the trajectories are determined, the resulting $r(t)$ and $\phi(t)$ are substituted into the following expressions:
\begin{equation}
x(t) = r(t) \cos[\phi(t)],\quad y(t) = r(t) \sin[\phi(t)],\quad z(t)=0
\end{equation}
From this, the second moment of the energy density $T^{00}/c^2$ is computed as \citep{Maggiore:2007ulw}
\begin{equation}
M_{ij} = \frac{1}{c^2}\int x_i x_j T^{00}(t, x) \, d^3x. 
\end{equation}
Note that for this orientation of the axes the binary system lies on the $xy$ plane, and we have $M_{13} = M_{23} = M_{33} = 0$. In the linearized theory's multipole expansion, which assumes weak fields and non-relativistic conditions, \(T_{00}/c^2\) can be replaced by the mass density to the lowest order in \(v/c\).
Using \( M_{ij} \), the plus and cross polarizations of the gravitational wave strains are computed as (for further details, refer to \citet{Maggiore:2007ulw})
\begin{align}\label{hplus}
h_+(t; \bar{\theta}, \bar{\phi}) &= \frac{G}{Rc^4} \left[ \ddot{M}_{11} \left(\cos^2 \bar{\phi} - \sin^2 \bar{\phi}\right) \cos^2 \bar{\theta} \right. \nonumber \\
&\quad\left. + \ddot{M}_{22} \left(\sin^2 \bar{\phi} - \cos^2 \bar{\phi}\right) \cos^2 \bar{\theta} - \ddot{M}_{12} \sin 2\bar{\phi} (1 + \cos^2 \bar{\theta}) \right]
\end{align}
\begin{align}\label{hx}
h_{\times}(t; \bar{\theta}, \bar{\phi}) &= \frac{G}{Rc^4} \left[ (\ddot{M}_{11} - \ddot{M}_{22}) \sin 2\bar{\phi} \cos \bar{\theta} + 2\ddot{M}_{12} \cos 2\bar{\phi} \cos \bar{\theta}\right]
\end{align}
where $R$ is the distance from the center of mass of the binary system to the observatory, and $\bar{\theta}$ and $\bar{\phi}$ describe the observer's position   in spherical coordinates relative to the center of mass of the binary system. 
The angle $\bar{\phi}$ specifies the observer's position within the orbital plane, indicating the azimuthal direction around the binary.
The angle $\bar{\theta}$ coincides with  the angle between the orbital plane's normal vector and the line of sight. If $\bar{\theta} = 0$, the observer is directly above the binary’s  positive $z-$axis. If $\bar{\theta} = \frac{\pi}{2}$, the observer is edge-on, viewing the binary from the plane of the orbit (from Equation~(\ref{hx}); in such a case, $h_{\times} = 0$). In all cases, the system is positioned at $\{R, \bar{\theta}, \bar{\phi}\} = \{1 \, \text{Mpc}, 0, 0\}$. This configuration follows \citet{Montalvo:2024iwq}, enabling a direct comparison between our results and theirs. Furthermore, the evaluation is performed at the retarded time \( t \), while the observed time is expressed as \( t_{\mathrm{obs}} = t + \frac{R}{c}. \)

\section{An Example of an EMRI Binary System in a Static Dark Matter Spike}
Consider an isolated binary system consisting of a supermassive black hole and a stellar-mass compact object, often referred to as an extreme mass ratio inspiral (EMRI), with a mass ratio (of the stellar-mass compact object to the supermassive black hole) in the range $10^{-5}$--$10^{-8}$ (see, for example, Table 9 in \citep{LISA:2022yao}). 

\subsection{A Model of an EMRI Binary System}
 In all cases, the two-body system consists of black holes, where \(m_1\) represents the mass of the central black hole and \(m_2\) denotes the mass of a smaller compact object (\(m_2 \ll m_1\)), a stellar-mass black hole  in a near-Keplerian orbit around the central black hole. 
The mass of the central black hole is set to \(m_1 = 10^6 \, \mathrm{M}_{\odot}\), while the mass of the orbiting black hole is \(m_2 = 10 \, \mathrm{M}_{\odot}\). When a medium is present, the smaller black hole acts as a perturber, moving through the medium and interacting with the gravitational wake it generates while orbiting the central compact object.

To focus on specific physical mechanisms, the orbit is consistently initialized to ensure a quasi-circular inspiral\endnote{In Figure~\ref{fig:0} we observed an extended phase during which $r$ decreases gradually, followed by a rapid plunge phase where the assumption of a quasi-circular orbit ceases to be valid \citep{Maggiore:2007ulw,Maggiore:2018sht}. The waveforms computed in Equations~(\ref{hplus}) and (\ref{hx}) remain valid only until the moment when the inspiral phase concludes, after which the two compact objects plunge toward each other and coalesce (see also \citep{Buonanno:2000ef}).}.
 Specifically, we assume $\dot{r}_{\rm init} = 0$ and choose $\dot{\phi}_{\rm init}$, such that it corresponds to the minimum of the effective potential when the 2.5~PN correction (or dynamical friction) is neglected.
In the following sections, we compute $\dot{\phi}_{\rm init}$ by including the 1~PN correction to the Newtonian gravitational field, as well as the global gravitational interaction between the medium and the perturber when a medium is considered.
 By incorporating 1~PN and 2.5~PN corrections through the solution of \mbox{Equations~(\ref{lagrangeeqr}) and (\ref{lagrangeeqphi})} the gravitational waveforms derived from Equations~(\ref{hplus}) and (\ref{hx}) have a period of \( T_{\rm GW} = 37.1410 \, \text{h} \), while the orbital phase, \( \phi(t) \), completes a cycle at around \( 2.0001 \times T_{\rm GW} \).

\subsection{A Static Dark Matter Spike Model}

A Schwarzschild black hole (BH) with mass \(m_1\) that grows adiabatically, forming a surrounding static dark matter (DM) profile, \(\rho_{\text{DM}}(r)\), from an initial Navarro--Frenk--White (NFW) profile \citep{Navarro:1996gj} is considered. During the BH's adiabatic growth, the DM halo contracts, forming a spike and significantly increasing the DM density near the BH horizon: 

\begin{equation}\label{DisFor}
\frac{d\mathbf{p}}{dt}\bigg|_{\text{DF}} = -\frac{4\pi G^2 m_2^2\rho_{\rm DM}(r)  }{v^3}\xi(v)\ln \Lambda \,\hat{\textbf{v}}
\end{equation}

\begin{figure}[H]%

\includegraphics[width =0.95 \columnwidth]{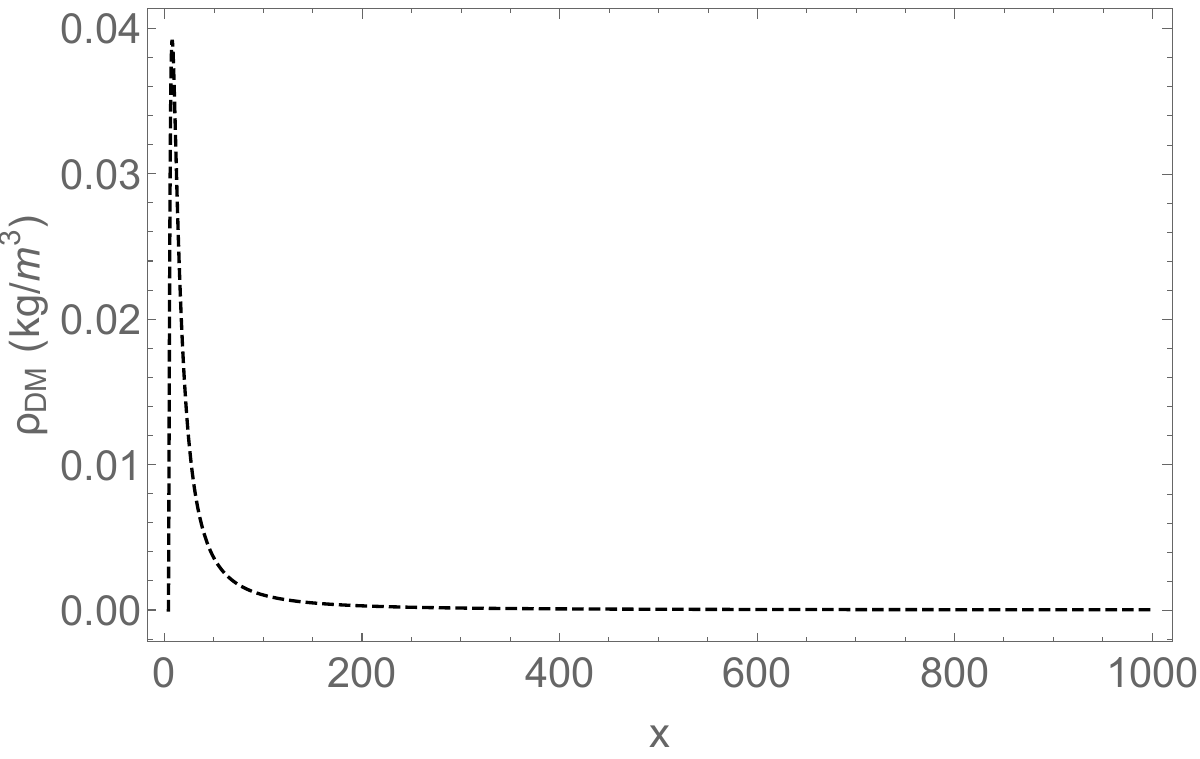}
\caption{A DM spike profile described by Equations~(\ref{dmprof}) and (\ref{dmconst}). Where $\alpha, \beta, \gamma,$ and $\delta$ are the relativistic NFW parameters (see \citet{Speeney:2022ryg} for details); $A, w,$ and $q$ are fit parameters with the following values: $\eta = 1$; $A = \frac{6.42 \times 10^{-43}}{1477.06^3} \times 1.989 \times 10^{30}$ (kg/m$^3$); $w = 1.82$; $q = 1.91$; $\tilde{\rho} = 0.5$ (GeV/cm$^3$); $m_1 = 10^6\rm M_{\odot}$; $a = 20$ (kpc); $\alpha = 0.331$; $\beta = -1.66$; $\gamma = 0.32$; $\delta = -0.000282$; and $x = \frac{c^2r}{G m_1}$.}
\label{fig:1}
\end{figure}

To model the relativistic DM density spike, we use the effective scaling function from Equation~(7) in \cite{Speeney:2022ryg}, given by
\begin{equation}\label{dmprof}
\rho_{\text{DM}}(r) = \tilde{\rho} 10^\delta \left(\frac{\rho_0}{0.3 \, \text{GeV/cm}^3}\right)^\alpha \left(\frac{m_1}{10^6 M_{\odot}}\right)^\beta \left(\frac{a}{20 \, \text{kpc}}\right)^\gamma, 
\end{equation}
with
\begin{equation}\label{dmconst}
\tilde{\rho} = A \left(1 - \frac{4}{x}\right)^w \left(\frac{4.17 \times 10^{11}}{x}\right)^q \,,
\end{equation}
\textls[-15]{where $x$ is defined as $x \equiv \frac{c^2 r}{G m_1}$. The DM halo is assumed to be static \citep{Kavanagh:2020cfn} and exhibits a DM spike profile with a scale density of \(\rho_0 = 0.5\) GeV/cm\(^3\) (see the caption of Figure~\ref{fig:1} \mbox{for details}).}

In the scenario of a collimated flow of collisionless particles, dynamical friction (DF) comes into play.  DF acts alongside gravitational wave emission \citep{Walker:1980zz,Pati:2000vt}, influencing the motion of the perturber \citep{Barausse:2007ph}.  
In the perturber's rest frame, the change in three-momentum due to the dynamical friction of the medium is expressed as
 \citep{Petrich:1988zz,Barausse:2007ph,Traykova:2021dua,Baumgarte:2024iby}: 

Here, \(\ln \Lambda\) represents the Coulomb logarithm, where \(\Lambda \equiv \frac{b_{\text{max}}}{b_{\text{min}}}\). We set \(b_{\text{max}}\) as the orbital radius of the perturber \citep{Hashimoto:2002yt,Fujii:2005kw,Kim_2007}, which naively represents the distance within which the gravitational influence of the background medium affects the perturber.
The capture impact parameter, \(b_{\text{min}}\), represents the effective size of the perturber during gravitational interaction with the medium. For a black hole, this is approximately given by \(b_{\text{min}} \approx 2G m_2 \left( 1 + \frac{v^2}{c^2} \right)/v^2\), where \(m_2\) denotes the mass of the stellar black hole (perturber) \citep{Petrich:1988zz,Barausse:2007ph}.
The perturber's velocity is indicated by \(v\), the Lorentz factor by \(\gamma(v) = \left[1 - \left(\frac{v}{c}\right)^2\right]^{-1/2}\), and \(\xi(v)=\gamma^2\left[1 + \left(\frac{v}{c}\right)^2\right]^2\) \citep{Traykova:2021dua,Vicente:2022ivh} accounts for the relativistic correction to the \mbox{dissipative force. }

The small object is subject to a dissipative force additional to gravitational wave emission, namely, dynamical friction from the DM halo. This results in an even faster inspiral, due to the loss of orbital energy. The dissipative force acts only parallel to \(\mathbf{v}\), as follows \citep{Montalvo:2024iwq}:
\begin{equation}
Q^{\text{DF}}_r =  -4\pi G^2 m_2^2 \frac{\xi(v) \rho_{\text{DM}}(r) \dot{r}}{(\dot{r}^2 + r^2 \dot{\phi}^2)^{3/2}} \ln (\Lambda), 
\end{equation}
\begin{equation}
Q^{\text{DF}}_{\phi}  = -4\pi G^2 m_2^2 \frac{\xi(v) \rho_{\text{DM}}(r) r^2 \dot{\phi}}{(\dot{r}^2 + r^2 \dot{\phi}^2)^{3/2}} \ln (\Lambda),
\end{equation}
 
The Lagrangian equations of motion, including post-Newtonian corrections, account for global gravitational interaction between the perturber and the dark matter (DM) spike within a  spherical shell of radius $r_{ 0}\equiv 4 r_{S}$, where $r_S$ is the Schwarzchild radius, to $r$ \citep{Speeney:2022ryg}, and local gravitational interaction via the dynamical friction on the perturber from the DM spike. These equations are given by
\begin{equation}\label{dm1}
    \ddot{r} - r\dot{\phi}^2 + \frac{Gm}{r^2}\left(1+\frac{m_{\rm DM}(r)}{m}\right)= Q^{\rm 1~PN}_{r} +Q^{\rm 2.5~PN}_{r}+Q_r^{\rm DF}  ,
\end{equation}
\begin{equation}\label{dm2}
    r^2\ddot{\phi} + 2r\dot{r}\dot{\phi} = Q^{\rm 1~PN}_{\phi}+Q^{\rm 2.5~PN}_{\phi}+Q_\phi^{\rm DF}    .
\end{equation}
where \(m_{\text{DM}}(r)=4\pi \int^r_{r_{\rm 0}}\rho(r')r'^2dr'\) denotes the mass of the dark matter (DM) spike profile enclosed within a spherical shell of radius \(r_{0}\) to $r$. 

\begin{figure}[H]%

\includegraphics[width =0.7\textwidth]{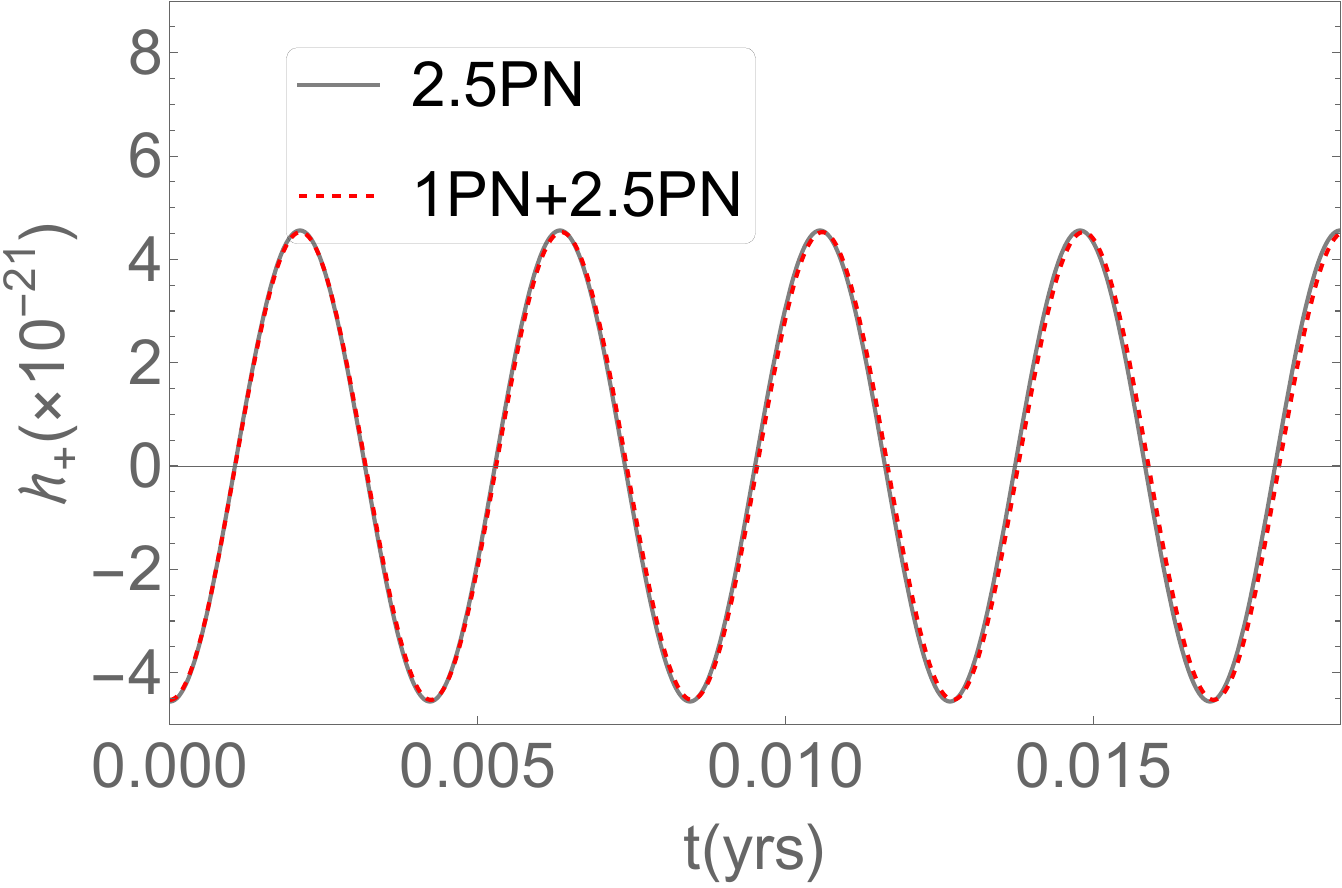}
\includegraphics[width =0.7\textwidth]{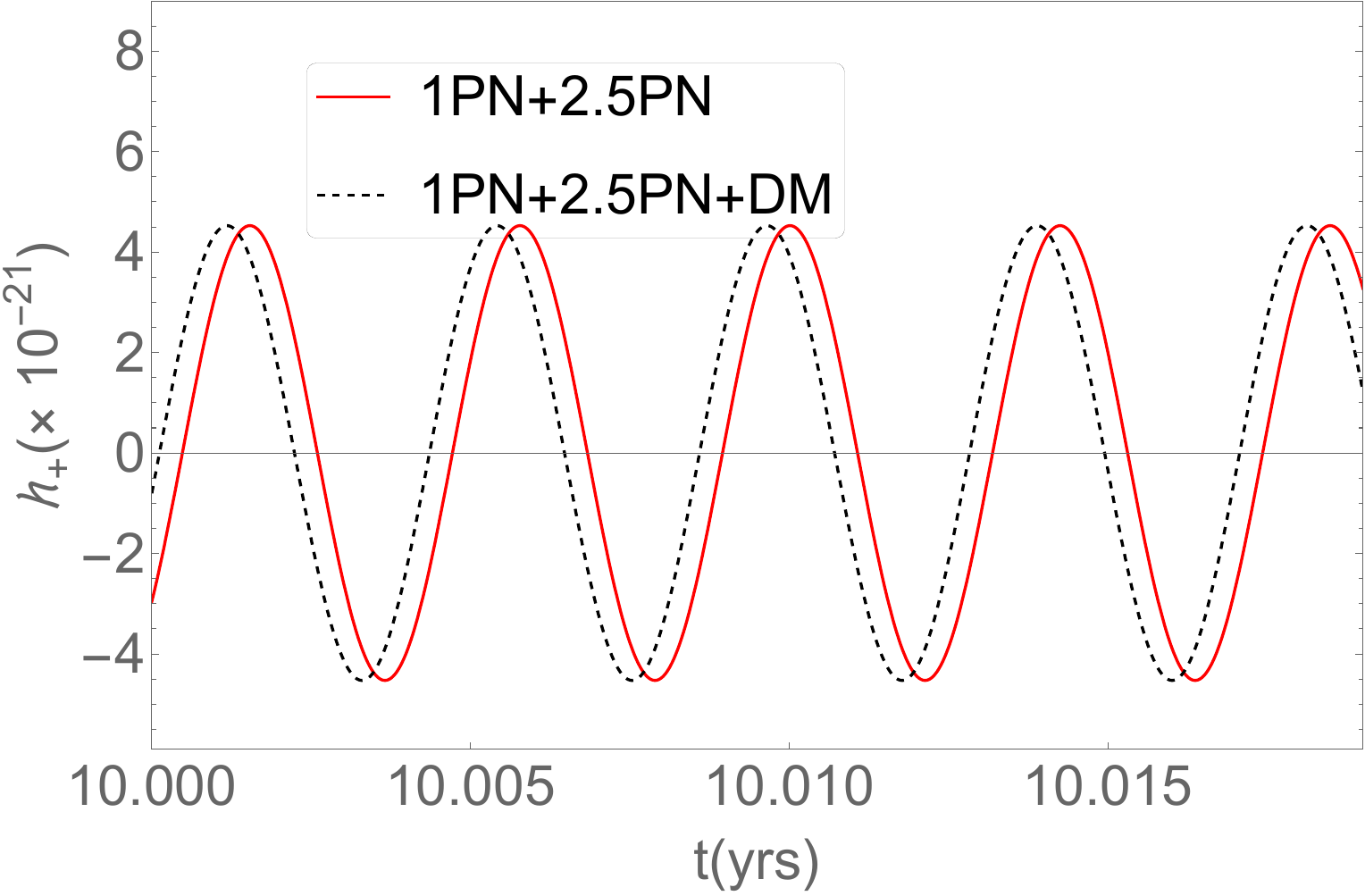}
\includegraphics[width =0.7\textwidth]{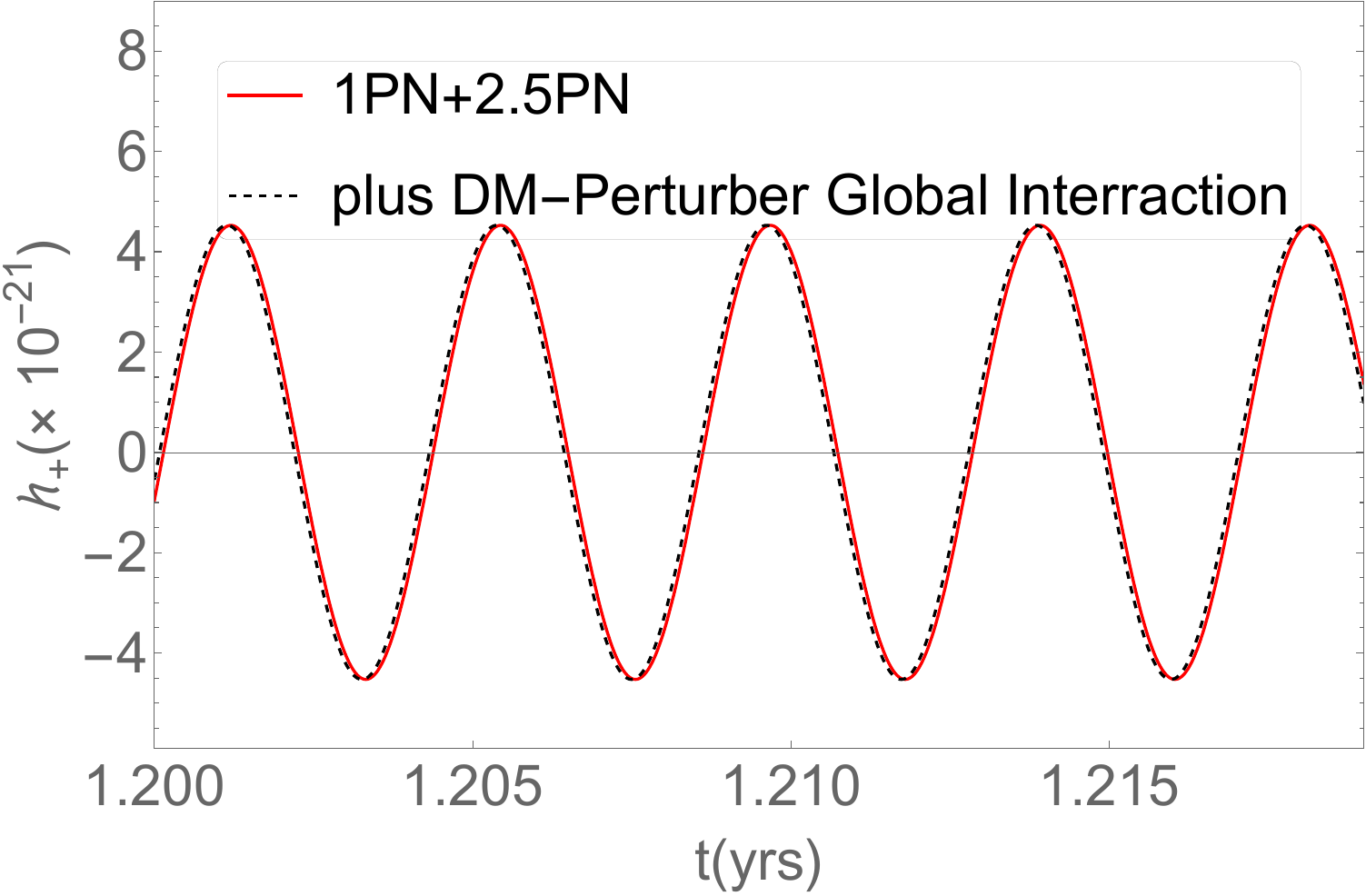}
\caption{The ``plus'' polarization amplitude, \( h_+(t) \), is plotted as a function of  time. Consider a binary system starting at a circular radius of $r_{\text{init}} = 70 r_{\text{ISCO}}$ at $t_{\rm init}=0$ and obtaining its orbit by solving the system of Equations~(\ref{dm1}) and (\ref{dm2}). 
 Upper Figure:   The  waveform evolution is depicted for the 2.5~PN correction (solid line) and for the combined 1~PN and 2.5~PN corrections (dashed line).
 Middle Figure:  The waveform evolution is shown for the combined 1~PN and 2.5~PN corrections (solid line) and for the 1~PN+2.5~PN corrections with a dark matter (DM) spike included (dashed line) through dynamical friction, in agreement with the results presented in Figure~3 of \citet{Montalvo:2024iwq}.
 Lower  Figure: The waveform evolution is depicted for 1~PN+2.5~PN+DM (dashed line)   including global gravitational effects from the dark matter (DM) spike, within a spherical shell from radius \( r_0 \) to the radius \( r \) of the perturber.  }
\label{fig:2}
\end{figure}

In Figure~\ref{fig:2}, we examine a binary system with an initial circular orbital radius of \( r_{\text{init}} = 70\, r_{\text{ISCO}} \), starting at \( t_{\text{init}} = 0 \). 
 The initial conditions are determined by incorporating the 1~PN correction (when it is assumed) and including the global radial force (when a dark matter spike is assumed) while neglecting friction terms. These conditions are established by setting \( \dot{r}_{\rm init} = 0 \) and determining \( \dot{\phi}_{\rm init} \), such that the orbit is initialized at the minimum of the potential (see, for example, Equation~(\ref{omegainit})).

 The  upper panel illustrates the `plus' polarization amplitude, \( h_+(t) \), over a time interval from 0  to 0.019 years. We present the waveform evolution for the  2.5~PN correction (solid line) and contrast this with the inclusion of  the 1~PN effect (dashed line). 

The middle panel shows the `plus' polarization amplitude, \( h_+(t) \), over a time interval from 10 years to 10.019 years. We adopt a constant value of \(\ln \Lambda \approx 3\) (see \citep{Chandrasekhar:1943ys, Eda:2014kra, Hannuksela:2019vip, Kavanagh:2020cfn, Speeney:2022ryg}) to align our results with those of the previous study \citep{Montalvo:2024iwq}. The waveform evolution is depicted for the 1~PN and 2.5~PN corrections (solid line) and compared with the inclusion of dynamical friction from the dark matter (DM) spike environment (dashed line), enabling a comparison and validating the results presented in Figure~3 of \citet{Montalvo:2024iwq}.

To zeroth order, angular momentum conservation implies that a smaller radius corresponds to a higher angular velocity $\dot{\phi}$. Dynamical friction opposes the orbital velocity but extracts energy and angular momentum, causing the body with mass $\mu$ to spiral inward more rapidly and thereby reach higher angular velocities sooner.
Note that the de-phasing is approximately $\Delta \phi \equiv \phi_{\rm DM} - \phi_{\rm No\,DM} \simeq 0.27\,\rm rad$ at 10 years, and it is positive. This indicates that the perturber delays in the absence of a DM spike, in contrast to the scenario where DM is present.

In the  lower panel, we further analyze the effects of the DM spike (dashed line) on the perturber, contrasting it with the case where the 1~PN+2.5~PN model is considered without a dark matter spike. Here, we also incorporate the global gravitational interaction between the perturber and the static DM spike.
 The DM spike is assumed to be confined within a spherical shell extending from the Schwarzschild radius \( r_0 \) to the radius \( r \) of the perturber.
This de-phasing effect, occurring at the 1-year mark, was not considered in the prior \mbox{study \citep{Montalvo:2024iwq},}  which neglected the global gravitational interaction with the DM spike, despite its dominance at such distances. Note that the $h_\times$ waveform is identical to $h_{+}$. 

\section{Steady-State Spherically Symmetric Energy Densities and Velocity Profiles of General Fluids in a Schwarzschild Black Hole Background}\label{accretingfluid}
In the previous section, the DM spike background was assumed to be static \citep{Montalvo:2024iwq}, and the effects of particle velocities were not considered. Here, we generalize the scenario of a black hole binary embedded in a static dark matter background by considering a general fluid in a steady state as it falls towards a Schwarzschild black hole of mass $m$.
 Stationarity assumes the black hole mass increases slowly, allowing the fluid distribution to adjust to the changing black hole metric over relevant space--time scales \citep{Babichev:2013vji}. The Schwarzschild metric  is given by \citep{Hobson:2006se}
\begin{equation}ds^2 = c^2 \left(1 - \frac{2Gm}{c^2r}\right) dt^2 - \left(1 - \frac{2Gm}{c^2r}\right)^{-1} dr^2 - r^2 d\theta^2 - r^2 \sin^2\theta d\phi^2.\end{equation}
The energy--momentum tensor of the perfect fluid is given by \citep{Hobson:2006se}
\begin{equation}
    T_{\mu\nu} = \left(\rho + \frac{ p}{c^2} \right) u_{\mu} u_{\nu} - p g_{\mu\nu}.
\end{equation}
Following \citet{Babichev:2004yx,Babichev:2005py,Babichev:2013vji}, we consider the accretion of a perfect fluid onto a Schwarzschild black hole. We consider a relativistic perfect fluid, disregarding the effects of viscosity and heat transport. Additionally, we assume that the fluid’s energy density is sufficiently low, such that its self-gravity can be neglected. The four-velocity, \( u^\mu \), is defined as follows:
\begin{equation}
    [u^\mu] = \left[\frac{dx^\mu}{d\tau}\right] = (u^t, u^{\rm r}, 0, 0)
\end{equation}
where $\tau$ represents the proper time. Owing to the spherical symmetry of the system, we have $u^\theta = 0$ and $u^\phi = 0$. This implies that all components of the four-velocity, along with the pressure $p \equiv p(\rho)$ and the mass energy density $\rho$, are functions of $r$ only.  Consequently, the four-velocity must satisfy the normalization condition \( u^\mu u_\mu = c^2 \), from which we \mbox{find that} 
\begin{equation}\label{norm}\frac{u^t}{c} = \sqrt{\frac{-\left(u^{\rm r}/c\right)^2g_{rr}  +1}{g_{tt}}}=\frac{\sqrt{1-\frac{2Gm}{c^2 r}+(u^{\rm r}/c)^2}}{1-\frac{2Gm}{c^2 r}}\end{equation}

Assuming local thermodynamic equilibrium of the fluid and that the flow is isentropic (which is straightforward to show; see \citep{Yang:2020bpj}), and applying the first law of thermodynamics (\(dE = -p \, dV\)) along with  \(V \equiv \frac{1}{n}\), where \(n\) denotes the effective fluid particle number density and \(V\) represents the specific volume per effective particle, we obtain the following expression \citep{Babichev:2004yx}:
\begin{equation*}
    d\left(\frac{\rho c^2}{n}\right)=-pd\left(\frac{1}{n}\right)\implies-p\frac{dn}{n^2}=c^2\frac{d\rho }{n}-\rho c^2\frac{dn}{n^2}
\end{equation*}
\begin{equation}\label{numberdens}
   \implies \frac{dn}{n} =\frac{d\rho}{\rho+\frac{p(\rho)}{c^2}}
\end{equation}
Equation~(\ref{numberdens}) implies that \begin{equation}\label{noverninfty0}\frac{n}{n_{\infty}} = \exp \left[ \int^{\rho}_{\rho_{\infty}} \frac{d\rho'}{\rho' + \frac{p(\rho')}{c^2}} \right]\end{equation}
where $n_{\infty},\rho_{\infty}$ represent the asymptotic values of the effective number density  and energy density, respectively.

The \( t \)-component of the energy--momentum conservation equation leads to the following relation:
\begin{equation}
 \nabla_{\mu} T^{t\mu} = 0 \implies   \frac{1}{\sqrt{|g|}}\partial_{\nu}\left(\sqrt{|g|}\,T^{\nu t}\right)+2\Gamma^{t}_{rt}T^{rt}=0\implies
\end{equation}
\begin{equation}\label{eqenergy}
   \frac{1}{r^2}\partial_{r}\left[r^2\left(\rho + p/c^2\right)u^tu^{\rm r}\right] +2\left(1-\frac{2Gm}{c^2r}\right)^{-1}\frac{Gm}{c^2r^2}(\rho+p/c^2)u^tu^{\rm r}= 0.
\end{equation} 
 By integrating Equation~(\ref{eqenergy}) and using Equation~(\ref{norm}), we arrive at another integral of motion, yielding the following expression \citep{Babichev:2004yx,Babichev:2013vji}:
\begin{equation}\label{integr2}
    (\rho+p/c^2)\left(1-\frac{2Gm}{c^2r}+\frac{\left(u^{\rm r}\right)^2}{c^2}\right)^{\frac{1}{2}}\left(\frac{c^2r}{Gm}\right)^2u^{\rm r}= \text{C}\equiv \rm const.
\end{equation}
It should be noted that a solution for $u^r = 0$ is only possible when $\text{C} = 0$. This becomes clearer by using the $r$-component of the energy--momentum conservation equation, which leads to
\begin{equation*}
    \nabla_{\mu} T^{r\mu} = 0 \implies
\end{equation*}
\begin{equation}\label{enrgmomentumf}
    x\left[1-\frac{2}{x}+\left(\frac{u^r}{c}\right)^2\right]\frac{dp}{dx}-x(u^r)^2\frac{d\rho}{dx}-\left(\rho c^2+p\right)\left[x^{-1}+2\frac{u^r}{c^2}\left(u^r+x\frac{du^r}{dx}\right)\right]=0
\end{equation}
where the dimensionless parameter, $x$, is defined by 
\begin{equation}\label{xdef}
x(r) \equiv \frac{c^2 r}{G m}.
\end{equation}

Consider a linear equation of state\endnote{The linear equation of state given by Equation~\eqref{EoS} provides a first-order approximation to any smooth curve \( p = p(\rho) \) \citep{Babichev:2004yx}. }, which describes scenarios such as a relativistic gas, phantom dark energy, and non-phantom dark energy, among others \citep{Babichev:2004qp, Babichev:2004yx}, \mbox{expressed as}
\begin{equation}\label{EoS}
  p(r) = \alpha\left[\rho(r) - \rho_0\right]c^2,
\end{equation}
where \(\rho_0\) and \(\alpha\) are parameters. We set \(\frac{p}{\rho c^2} = w\) to relate Equation~(\ref{EoS}), which is defined in terms of \(\alpha\) and \(\rho_0\), to the equation of state parameter \(w \equiv w(r)\), as shown below \citep{Babichev:2005py}:

\begin{equation}\label{wpar}
    w(r) = \alpha \frac{\rho(r) - \rho_0}{\rho(r)}.
\end{equation}

The sound speed of the fluid, $c_{\rm s}$, is defined as $c_{\rm s}^2 \equiv \frac{\partial p}{\partial \rho}$. A fluid is considered \textit{stable} if $0 < c_{\rm s}^2 \leq c^2$ and \textit{unstable} if $c_{\rm s}^2 > c^2$ or $c_{\rm s}^2 < 0$ \citep{Babichev:2004qp, Babichev:2004yx}.
  Therefore, for a  stable fluid, the squared effective speed of sound is defined as \( 0 < c_{\rm s}^2 = \frac{\partial p}{\partial \rho} = \alpha c^2\leq c^2 \), which implies that \( 0 < \alpha \leq 1 \).

\subsection{Static  Fluid in a Schwarzschild Background}

 From Equation~(\ref{enrgmomentumf}), in order to obtain a static fluid, we impose the condition that $u^r = 0$ and $\frac{du^r}{dx} = 0$ for all $x>2$.  Therefore, Equation~(\ref{enrgmomentumf}) implies hydrostatic equilibrium:
\begin{equation}\label{staticenergy}
     \frac{dp}{dx} = -\frac{\rho c^2 + p}{x^2\left( 1 - \frac{2}{x} \right)},
\end{equation}
From Equation~(\ref{staticenergy}), it is evident that \( p = -\rho c^2 \) if and only if \( \frac{dp}{dx} = 0 \). {However, the derivative \( \frac{dp}{dx} \) in Equation~(\ref{staticenergy}) diverges at \( x = 0 \) and \( x = 2 \). This naturally follows from the fact that \( x = 0 \) corresponds to the curvature singularity of the Schwarzschild black hole, while \( x = 2 \) marks the event horizon.}


From Equation~(\ref{enrgmomentumf}), it follows that for a linear equation of state, Equation~(\ref{EoS}), the only static and stable fluid that remains non-singular at $x=2$, in the sense that \( \rho(x) \) remains finite at $x=2$, in the exterior of a Schwarzschild black hole is characterized by \(w = -1\), corresponding to a cosmological constant.
By separating the variables in Equation~(\ref{staticenergy}) and integrating, \mbox{we obtain}
\begin{equation}\label{stat1}
    \int c_{\rm s}^2 \frac{d\rho}{\rho c^2 + p} = -\int \frac{dx}{x^2 \left( 1 - \frac{2}{x} \right)}.
\end{equation} 
 By substituting the equation of state (EoS) from Equation~(\ref{EoS}) into Equation~(\ref{stat1}), a solution can be obtained for $\alpha \neq 0, -1$ and $\rho_0 \neq 0$, as
\begin{equation}
    \rho(x)=\frac{\alpha}{1+\alpha}\rho_0+B \left(\frac{x}{2-x}\right)^{\frac{1+\alpha}{2 \alpha}}
\end{equation}
where $B$ is an integration constant. For a stable fluid, $\rho(x)$ diverges at $x = 2$, unless $B = 0$.
 If $B=0$, the density simplifies to the constant value $\rho=\rho_{\Lambda} \equiv \frac{\alpha}{1+\alpha} \rho_0$, and by substituting into Equation~(\ref{wpar}) it follows that $w = -1$. In the following subsection, it can be seen that the accreting profiles $\rho(x)$ of stable fluids also remain finite at $x = 2$.

\subsection{Accreting   Fluid in a Schwarzschild Background}

For any fluid in a gravitational field, the condition \( u^\mu u_\mu = c^2 \) implies that \( u_{\mu}\nabla_{\nu}u^{\mu} = 0 \). Using this, the equation 
\( u_\nu \nabla_{\mu} T^{\mu\nu} = 0 \) leads to the following equation \citep{Hobson:2006se}:
\begin{equation}\label{cont}
    \nabla_\mu \left( \rho u^\mu \right) + \frac{p}{c^2} \nabla_\mu u^\mu = 0,
\end{equation}
 where $\rho, p$ are the rest frame density and the pressure of
the fluid. From Equation~(\ref{cont}), \mbox{we obtain}
\begin{equation}\label{eqfluid}
    u^{\rm r} \partial_r \rho+\frac{1}{\sqrt{|g|}}(\rho+p/c^2)\partial_{\mu}\left(\sqrt{|g|}\, u^{\mu}\right)=0
\end{equation}
From the integration of Equation~(\ref{eqfluid}), and given that \( u^{r} \neq 0 \),
\begin{equation}\label{integr1}
    \int\frac{d\rho}{\rho+p/c^2}+\int\frac{2}{r}dr+\int\frac{du^{\rm r}}{u^{\rm r}}=0
\end{equation}
 Given Equation~(\ref{numberdens}), then Equation~(\ref{integr1}) gives an integral of motion \citep{Babichev:2013vji}
\begin{equation}\label{eqint01}
    \frac{u^{\rm r}}{c} \left(\frac{c^2r}{Gm}\right)^{2}\frac{n}{n_{\infty}} = -A, 
\end{equation}
where $A$ is a positive dimensionless constant, with \( u^{\rm r} \equiv \frac{dr}{d\tau} < 0 \) indicating direction towards the center (accretion).  {From  Equation~(\ref{integr2}), and given Equation~(\ref{eqint01}), it follows that
\begin{equation}\label{integr2'}
    \frac{\rho+p/c^2}{n}\left(1-\frac{2Gm}{c^2r}+\frac{\left(u^{\rm r}\right)^2}{c^2}\right)^{\frac{1}{2}}=-\frac{\text{C}}{n_{\infty}A}\equiv  \frac{ \rho_{\infty}+ p_{\infty}/c^2}{n_{\infty}}
\end{equation}
 where $\rho_{\infty},p_{\infty}$ are the energy density and pressure at infinity. Note that the constant ratio $-C/A$ is determined by taking the limit as \(r \to \infty\).}

{Furthermore, by integrating the flux of the fluid over the two-dimensional surface of the black hole (we are assuming that the BH is not moving \citep{Babichev:2005py}), we obtain  $\dot{m} = - \frac{ G^2}{c^3} \int T_{t}^{\, r} \sqrt{-g} \, d\theta \, d\phi=\frac{-4\pi G^2}{c^3}r^2T^{\,r}_{t}$. By combining the above equation with Equations~(\ref{eqint01}) and (\ref{integr2'}), the following relationship is obtained \citep{Babichev:2004yx,Babichev:2013vji,Debnath:2015hea,Bahamonde:2015uwa}:\vspace{-4pt}
\begin{equation}\label{dem0}
    \frac{\dot{m}}{m^2} = \frac{ 4\pi A G^2}{c^3} [\rho_{\infty} + p_{\infty}/c^2].
\end{equation}
This result is valid for any equation of state \( p = p(\rho) \). During the accretion process, the rate of change of the total energy inside the black hole evolves slowly. The accretion remains self-consistent as long as the accreting fluid is light and the black hole mass increases sufficiently slowly (details can be found in \citep{Babichev:2013vji}). }

 Using the definition of the dimensionless parameter in Equation~(\ref{xdef})
 and obtaining the differentials of Equations~(\ref{eqint01}) and (\ref{integr2'}), we obtain \citep{Michel:1972oeq}\vspace{-4pt}
\begin{equation}\label{brack}
   \frac{du^{\rm r}}{u^{\rm r}} \left[ V_{\rm c}^2 - \frac{(u^{\rm r})^2}{ 1 - \frac{2}{x} + \left(u^{\rm r} / c \right)^2 } \right] + \frac{dx}{x} \left[ 2V_{\rm c}^2 - \frac{c^2}{x \left( 1 - \frac{2}{x} + \left(u^{\rm r} / c \right)^2 \right)} \right] = 0 
\end{equation}
where we denote\vspace{-4pt}
\begin{equation}\label{vfluid}
    V_c^2 \equiv c^2 \left[ \frac{n}{\rho + p / c^2} \frac{d(\rho + p / c^2)}{dn} - 1 \right].
\end{equation}
   From Equation~(\ref{numberdens}), it is easy to show that $V_{\rm c}$ has units of velocity and coincides with the sound speed of the fluid ($V_{\rm c}^2=c_{\rm s}^2$). Then, for  stable fluids with $0<c^2_{\rm s} < c^2$ the constant \( A \) is determined by the condition that the fluid flow transitions smoothly through the critical point (or sonic point), i.e.,  the point where  the fluid's velocity equals the local speed of sound (and marks the transition from subsonic flow to supersonic flow, which takes place at the sonic sphere located at $r=r_{\rm c}$) \citep{Michel:1972oeq,Babichev:2004yx}.
  
   {An observer in the rest frame of the fluid would measure the local sound speed $V_{\rm c}$.    If the accreting fluid is perturbed at $P$, and $P$ lies on the sonic sphere, where the radial velocity of the accreting fluid, as seen by a stationary observer at $P$, matches the local sound speed, then a wavefront moving radially outward would appear stationary to the stationary observer at $P$. Note that the ordinary radial velocity of the fluid for an observer stationary at a point $P$  is calculated as (using Equation~(\ref{norm})) $\frac{dr}{dt}\big|_{P}=c(1-2/x)^{-1}\frac{u^r}{u^t} = \frac{u^r}{\sqrt{1 - 2/x + \left(u^{\rm r} / c \right)^2}}$. This means that at $x = x_{\rm c}$, the condition $V_{\rm c}^2 - ( u_{\rm c}^{r} )^2  [1 - 2/x_{\rm c} + ( u_{\rm c}^{r}/c )^2]^{-1} = 0$ holds.}
   
    Suppose that a critical point exists such that $2 \leq x_{\rm c} < \infty$ and that Equation~(\ref{brack}) can be rewritten as
\begin{equation}\label{durdx}
       \frac{du^r}{dx}=-2\frac{u^r}{x}\,\frac{  \frac{V_{\rm c}^2}{c^2}-\frac{1}{2x \left( 1 - 2/x + \left(u^{\rm r} / c \right)^2 \right)} }{ \frac{V_{\rm c}^2}{c^2} - \frac{(u^{\rm r})^2/c^2}{ 1 - 2/x + \left(u^{\rm r} / c \right)^2 } }
   \end{equation}

   Assuming at the same time that $V_{\rm c}^2/c^2 - (2x_{\rm c})^{-1}[ 1 - 2/x_{\rm c} + ( u_{\rm c}^{r}/c )^2 ]^{-1} \neq 0$, it follows that $\frac{du^{r}}{dx}\big|_{x_{\rm c}} = \infty$ (see Equation~(\ref{durdx})), which implies that $u^{r}$ is not sufficiently differentiable at that point.
Based on the physical argument that the fluid accelerates toward the black hole with a regular flow throughout, we expect $0 < \frac{du^r}{dx} < \infty$ at every point for $x \geq 2$. Consequently, by the definition of the derivative, the left-hand and right-hand limits of Equation~(\ref{durdx}) must converge to the same definite value.

Such a regular flow should also occur at the critical point, implying that $\lim_{x \to x_{\rm c}} \frac{du^r}{dx} = \frac{0}{0}$, which requires both terms in Equation~(\ref{brack}) to vanish simultaneously.
  This may allow a smooth ``transition'' of the flow through the critical point \citep{Michel:1972oeq},  leading to  the following relations, which are used to determine $A$ \citep{Babichev:2004yx}:\vspace{-6pt}
\begin{equation}\label{critical}
   \left( u_{\rm c}^r\right)^2=\frac{c^2}{2x_{\rm c}},\quad V^2_{\rm c}=\frac{\left(u_{\rm c}^r\right)^2}{1-3\left(u_{\rm c}^r/c\right)^2}
\end{equation}
Given Equations~(\ref{critical}) and (\ref{EoS}), with \(V_{\rm c}^2 = \partial p / \partial \rho\), the critical values are computed as\vspace{-6pt}
\begin{equation}\label{criticalvalues}
        x_{\rm c}=\frac{1+3\alpha}{2\alpha},\quad \left(\frac{u_{\rm c}^{r}}{c}\right)^2=\frac{\alpha}{1+3\alpha}.
    \end{equation}

The effective number density ratio \(n/n_{\infty}\) is obtained by integrating Equation~(\ref{noverninfty0}), given the EoS in Equation~(\ref{EoS}):\vspace{-6pt}
    \begin{equation}\label{noverninf}
        \frac{n(\rho)}{n_{\infty}}=\left(\frac{\rho+p(\rho)/c^2}{\rho_{\infty}+p_{\infty}/c^2}\right)^{\frac{1}{1+\alpha}}.
    \end{equation}

\textls[-25]{The radial four-velocity $u^{\rm r}$ of the fluid flow and energy density as functions of radius, $\rho(r)$, are determined  using the following set of equations, obtained from Equations~(\ref{eqint01}), (\ref{integr2'}), and (\ref{noverninf}):}\vspace{-3pt}
\begin{equation}\label{integr1lin}
    1-2/x + \left(\frac{u^{\rm r}}{c}\right)^2 = \left(-\frac{u^{\rm r} x^2}{c\,A}\right)^{2\alpha},
\end{equation}
\begin{equation}\label{integr2lin}
    \frac{\rho + p/c^2}{\rho_{\infty} + p_{\infty}/c^2} = \left(-\frac{c\,A}{u^{\rm r} x^2}\right)^{1+\alpha}.
\end{equation}
For $0 < \alpha \leq 1$, the constant \( A \), which determines the flux onto the Schwarzschild black hole, is given by Equations~(\ref{integr1lin}) at the critical point and (\ref{criticalvalues}), as described in \citep{Babichev:2004yx}:\vspace{-4pt}
\begin{equation}\label{Aconstant}
    A = \frac{(1 + 3\alpha)^{\frac{1+3\alpha}{2\alpha}}}{4\alpha^{\frac{3}{2}}}.
\end{equation}

\begin{figure}[H]%

\includegraphics[width = 0.82\textwidth]{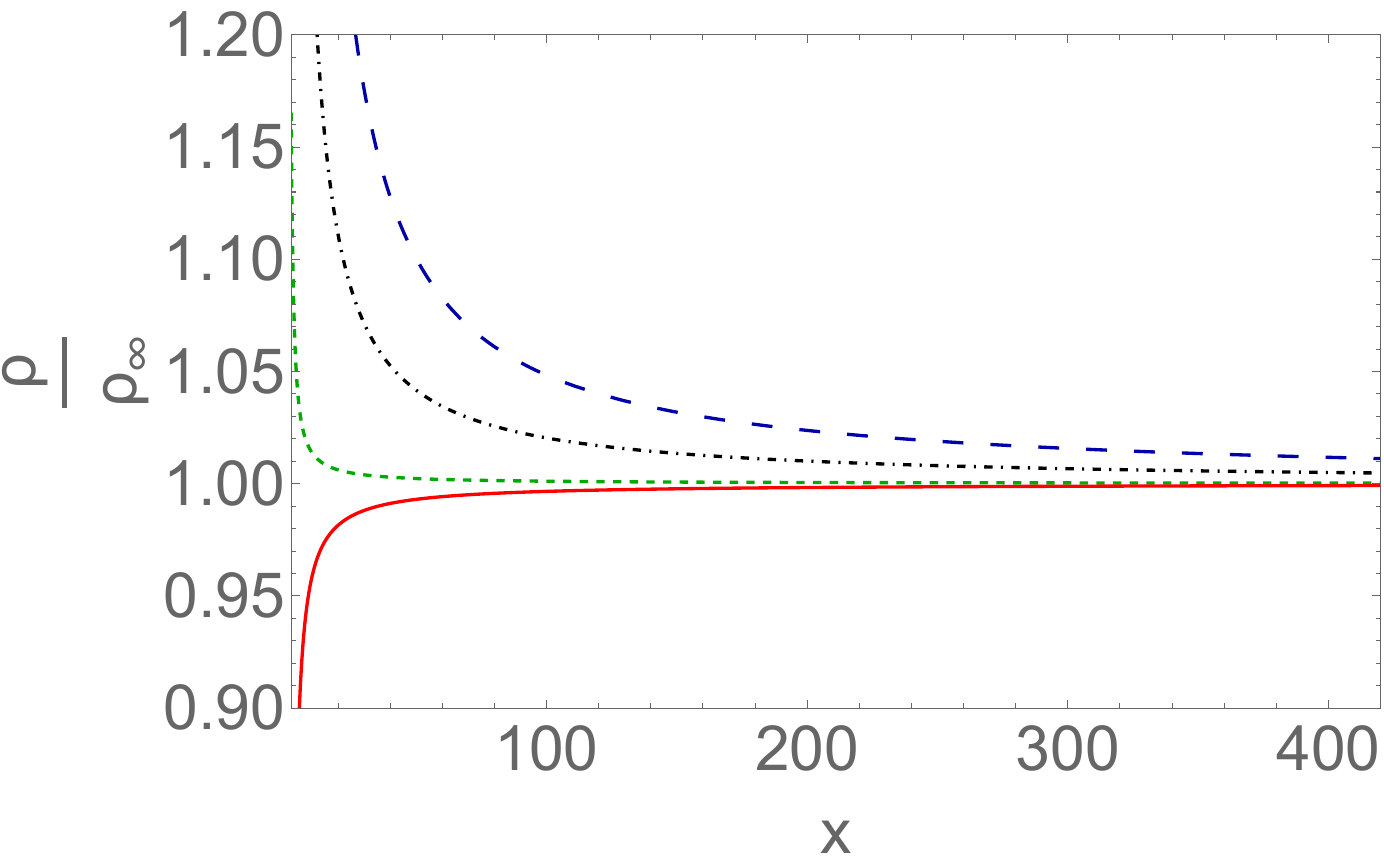}
\includegraphics[width = 0.82\textwidth]{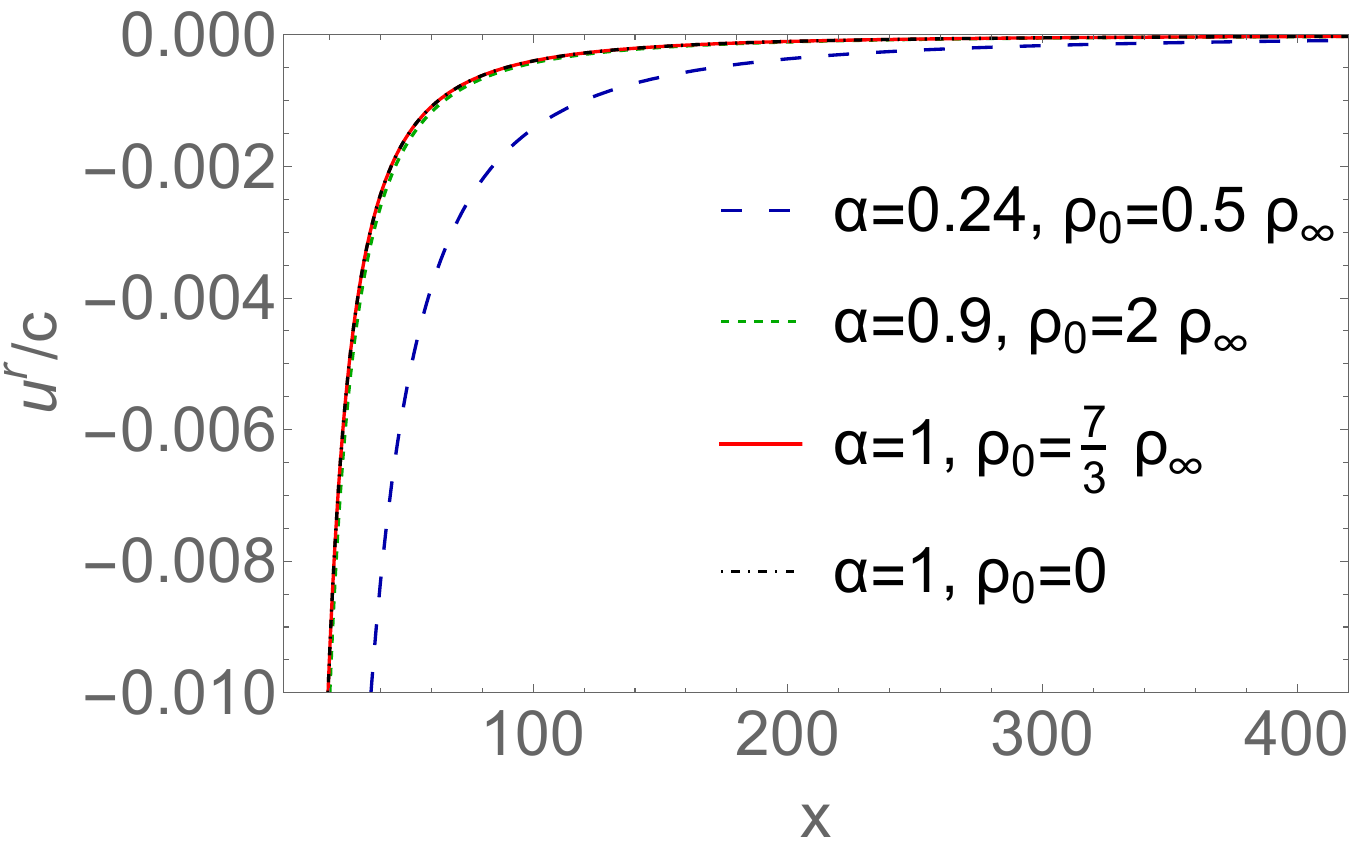}
\includegraphics[width = 0.82\textwidth]{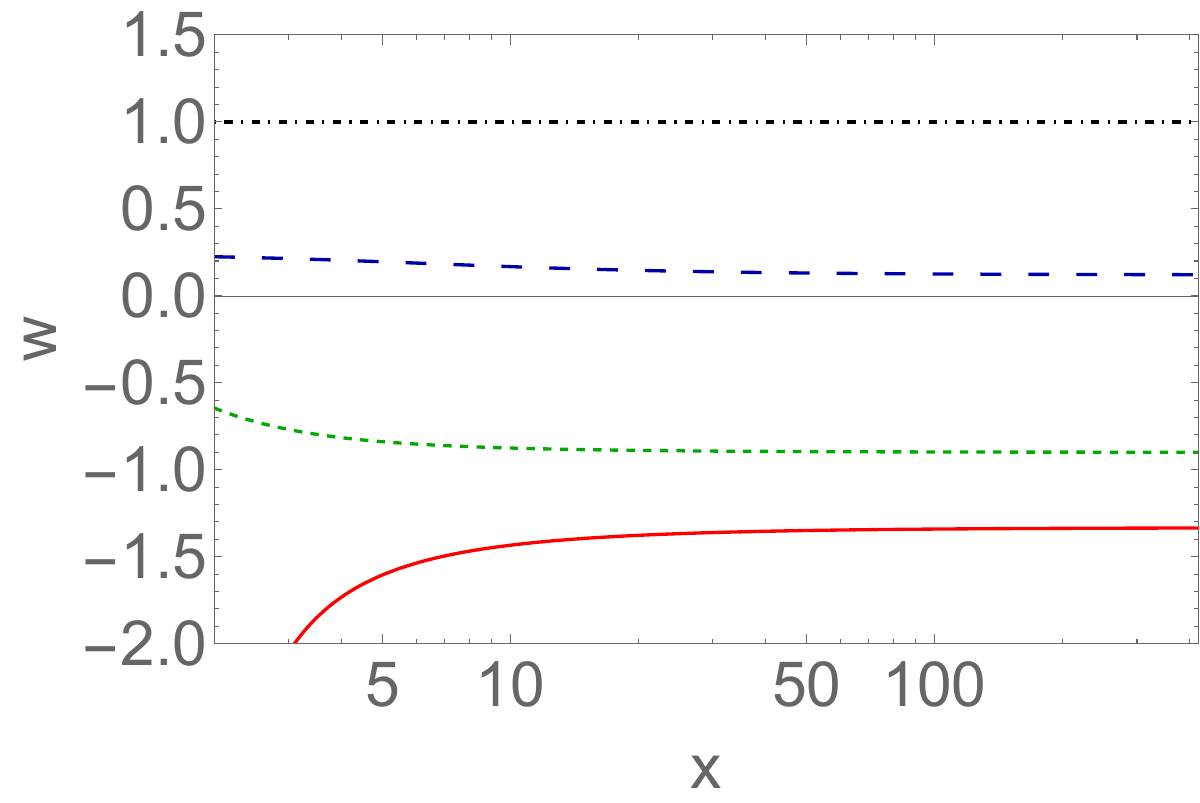}
\caption{Several stable fluid models accrete onto a Schwarzschild black hole, each obtained by solving the set of Equations~(\ref{integr1lin}) and (\ref{integr2lin}) and determining the constant \(A\), using Equation~(\ref{Aconstant}). All models are presented in terms of the dimensionless variable \(x \equiv \frac{c^2 r}{G m}\). For each model, the energy density ratio \(\frac{\rho}{\rho_{\infty}}\), the radial velocity component \(u^{r}/c\) (as a fraction of the speed of light), and the corresponding EoS parameter \(w\) are plotted as functions of \(x\), up to \(x = 420\), which corresponds to \(70\, r_{\mathrm{ISCO}}\).}
\label{fig:3}
\end{figure}

 Note that specific values of the parameter \(w_{\infty}=\alpha(1-\rho_0/\rho_{\infty})\) are obtained for different \((\alpha, \rho_0)\) pairs, with \(w_{\infty}\) denoting the EoS parameter at infinity. As can be deduced from Equation~(\ref{integr1lin}) and Figure~\ref{fig:3}, the radial velocity profile is solely determined by the parameter \(\alpha\), with smaller \(\alpha\) values leading to higher absolute velocities. After deriving the velocity profile, the corresponding energy density profile is calculated through Equation~(\ref{integr2lin}), showing that smaller \(\alpha\) values result in higher absolute density values.

In Figure~\ref{fig:3}, we examine several stable fluid models, including the energy density ratio \(\frac{\rho}{\rho_{\infty}}\), the radial velocity component \(u^{r}/c\) (as a fraction of the speed of light), and the corresponding EoS parameter \(w\), in terms of the dimensionless variable $x$, by solving the set of Equations~(\ref{integr1lin}) and (\ref{integr2lin}) and determining the constant \(A\) using Equation~(\ref{Aconstant}).
For instance, in a model with \(\alpha = 0.24\) and \(\rho_0 = 0.5 \rho_{\infty}\), the equation of state parameter is \(w(x=420) \simeq 0.12\), while for \(\alpha = 1\) and \(\rho_0 = 0\) we have \(w = 1\). Additionally, we consider linear models of dark energy with the following parameters: \(\alpha = 0.9\), \(\rho_0 = 2 \rho_{\infty}\), and \(w(x=420) \simeq -0.90\) (quintessence); and \(\alpha = 1\), \(\rho_0 = \frac{7}{3} \rho_{\infty}\), with \(w(x=420) \simeq -1.34\) (phantom).
 As previously suggested \citep{Babichev:2004yx}, quintessence dark energy tends to form an overdensity around black holes, whereas phantom dark energy leads to an underdensity\endnote{Notably, as indicated by Equation~(\ref{dem0}), the accretion of a phantom test fluid onto a Schwarzschild black hole can result in a reduction of the black hole's mass~\citep{Babichev:2004yx}. For further discussions in the context of cosmological expansion, see~\mbox{\citet{Gao:2008jv},} \citet{Martin-Moruno:2009cmc,Karkowski:2012vt}.}.

\section{Equations of Motion for an EMRI Binary System Within an Accreting Dark Fluid} 
 As long as \( m_1 \ll m_2 \), such that the center of mass nearly coincides with the center of the supermassive black hole, we can approximate \( \mu \simeq m_2 \). Under this assumption, the two-body system can be treated as a perturber of mass \( m_2 \), initially in a circular orbit and sufficiently far from a Schwarzschild black hole of mass \( m_1 \), which is surrounded by a fluid at rest at infinity. 
The Lagrangian equations of motion, which include the post-Newtonian corrections  as generalized forces and the gravitational interactions between the perturber and the dark fluid, are as follows:
\begin{equation}\label{lagrangeeqr}
    \ddot{r} - r\dot{\phi}^2 + \frac{Gm}{r^2}-f^{\rm fluid}_r= Q^{\rm 1~PN}_{r} +Q^{\rm 2.5~PN}_{r}  ,
\end{equation}
\begin{equation}\label{lagrangeeqphi}
    r^2\ddot{\phi} + 2r\dot{r}\dot{\phi} = Q^{\rm 1~PN}_{\phi}+Q^{\rm 2.5~PN}_{\phi}   .
\end{equation}
where $\mathbf{f}^{\rm fluid}_r=-\nabla \Phi$, the gravitational interaction between the perturber and the dark fluid 
is enclosed within a spherical shell of radius \( r_S \) to \( r \), and $\Phi$ is the gravitational potential obtained through $\nabla^2 \Phi = 4\pi G \left(\rho + \frac{3p}{c^2}\right)$. We compute the gravitational force numerically by solving the Poisson equation. This force is evaluated through numerical integration and is explicitly added as a generalized force in Equation~(\ref{lagrangeeqr}):
\begin{equation}\label{frfluid}
    Q^{\rm global}_{r} =f_r^{\rm fluid} = -4\pi Gr^{-2} \int_{r_S}^{r} \left(\rho(r) + 3\frac{p(r)}{c^2}\right)r^2 \, dr,
\end{equation}
In the case of a static fluid \(\left( u^r = 0 \right)\) with \(w = -1\) (cosmological constant),   the presence of the fluid can be detected only through the gravitational interaction via Equation~(\ref{frfluid}) on the perturber. We initiate the orbit with the re-scaled initial angular velocity, which is determined after incorporating the generalized force $Q_r^{\rm global}$, and is given by
\begin{equation}
\phi'_{\text{init}} =\frac{\sqrt{  2 (2 +\eta) e^2 +( f\,\bar{r}_{\rm init}^2-1)\bar{r}_{\rm init}}}{\bar{r}_{\rm init}^{3/2} \sqrt{ e^2 (1+3\eta)- \bar{r}_{\rm init}}}
\end{equation}
where $f\equiv \frac{t_{\rm yr}}{c} f_r^{\rm fluid}$. Lastly, the quadrupole moment tensor in such a case will be
\begin{equation}\label{quadropole}
M_{ij} =\frac{1}{c^2} \int x_i x_j \left[\mu c^2 \delta\left(x-x(t)\right)\delta\left(y-y(t)\right)\delta(z)+T^{00}_{\rm fluid}\right] \, d^3x. 
\end{equation}
The energy density of the fluid, as represented in the stress--energy tensor, is \( T^{00}_{\text{fluid}}(r) = \left(\rho(r)+p(r)/c^2\right)\left[u^t(r)\right]^2-p(r)\left(1-\frac{2Gm}{c^2r}\right)^{-1}\).

\subsection{Effects of Dynamical Friction}
In this framework, we model the local interaction between the perturber and the fluid as the perturber traversing a non-self-gravitating, collisional fluid.\endnote{This is not to suggest that dark energy, for example, acts as such a fluid. Nonetheless, this model offers valuable qualitative insights.}
The tangential and radial components of the three-force due to the gravitational interaction of a perturber with a non-self-gravitating, collisional fluid, where the perturber is moving at relativistic velocities with respect to the fluid, as observed by a stationary observer located far from the central object, are given by\endnote{For simplicity, we assume that the medium is spherically symmetric around the perturber \citep{Barausse:2007ph}.}
 \citep{Barausse:2007ph}
\begin{equation}\label{dpdtr}
   F^r=-\frac{4\pi (\rho + \frac{p}{c^2}) G^2 m_2^2 }{\tilde{V}^2}\xi(\tilde{V}) I_r ,
\end{equation}
\begin{equation}\label{dpdtphi}
   F^{\phi}  =-\frac{4\pi (\rho + \frac{p}{c^2}) G^2 m_2^2 }{r \, \tilde{V}^2}\xi(\tilde{V}) I_\phi ,  
\end{equation}
where $\rho$ and $p$ represent the rest frame density and the pressure of the fluid, respectively, and $\tilde{V}$ denotes the velocity of the perturber relative to the fluid flow (see Appendix \ref{AppendixA}).

The terms $I_r$ and $I_\phi$ (the orbital decay is attributed  to the azimuthal drag \citep{Kim_2007}) have been computed from \citet{Kim_2007} as (see also \citet{Barausse:2007ph})
\begin{equation}\label{ir}
I_r = 
\begin{cases}
M^2 10^{3.51M-4.22}, & \text{for } M < 1.1, \\
0.5 \ln \left[9.33M^2\left(M^2 - 0.95\right)\right], & \text{for } 1.1 \leq M < 4.4, \\
0.3 M^2, & \text{for } M \geq 4.4,
\end{cases}
\end{equation}
and
\begin{equation}\label{iphi}
I_\phi = 
\begin{cases}
0.7706 \ln\left(\frac{1+M}{1.0004 - 0.9185M}\right) - 1.4703M, & \text{for } M < 1.0, \\
\ln\left[330\Lambda(M - 0.71)^{5.72}M^{- 9.58}\right], & \text{for } 1.0 \leq M < 4.4, \\
\ln\left[\frac{\Lambda}{0.11M + 1.65}\right], & \text{for } M \geq 4.4,
\end{cases}
\end{equation}
where $M$ the Mach number of the perturber, $M=\tilde{V}/V_{\rm c}$ \citep{Barausse:2007ph}. These fits are accurate within 4\% for $M < 4.4$ and within 16\% for $M > 4.4$ \citep{Kim_2007,Barausse:2014tra}.
Note that \(\Lambda \equiv \frac{b_{\text{max}}}{b_{\text{min}}}\), as defined earlier, where the minimum impact parameter for a perturber, which is a black hole, is approximately given by \(b_{\text{min}} \approx \frac{2G m_2}{\tilde{V}^2} \left( 1 + \frac{\tilde{V}^2}{c^2} \right)\) \citep{Barausse:2007ph}.
The corresponding generalized forces resulting from friction in the radial and angular directions are given by \( Q^{\rm DF}_{r} = F_{r} \) and \( Q^{\rm DF}_{\phi} = rF_{\phi} \), as derived by combining Equations~(\ref{dpdtr})--(\ref{iphi}). These forces are subsequently incorporated into Equations~(\ref{lagrangeeqr}) and (\ref{lagrangeeqphi}).

\begin{figure}[H]%

\includegraphics[width =0.9\textwidth]{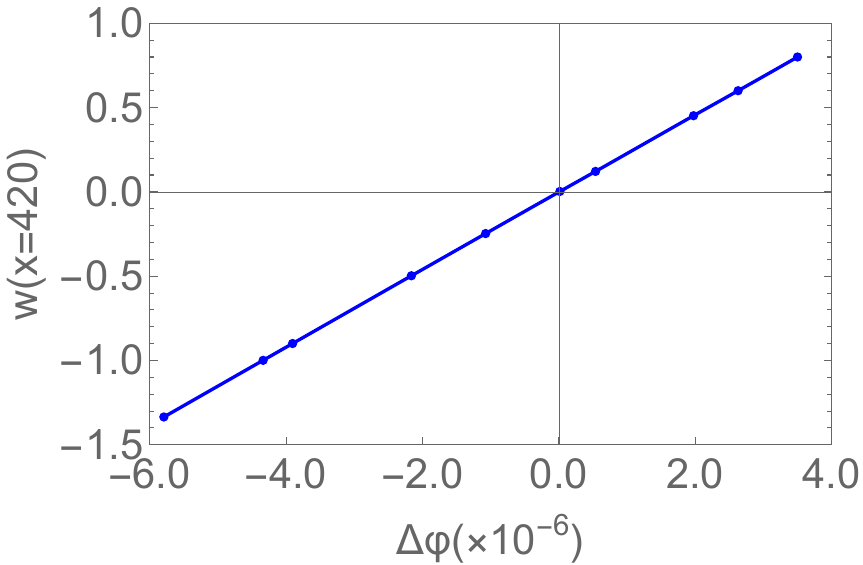}
\includegraphics[width =0.9\textwidth]{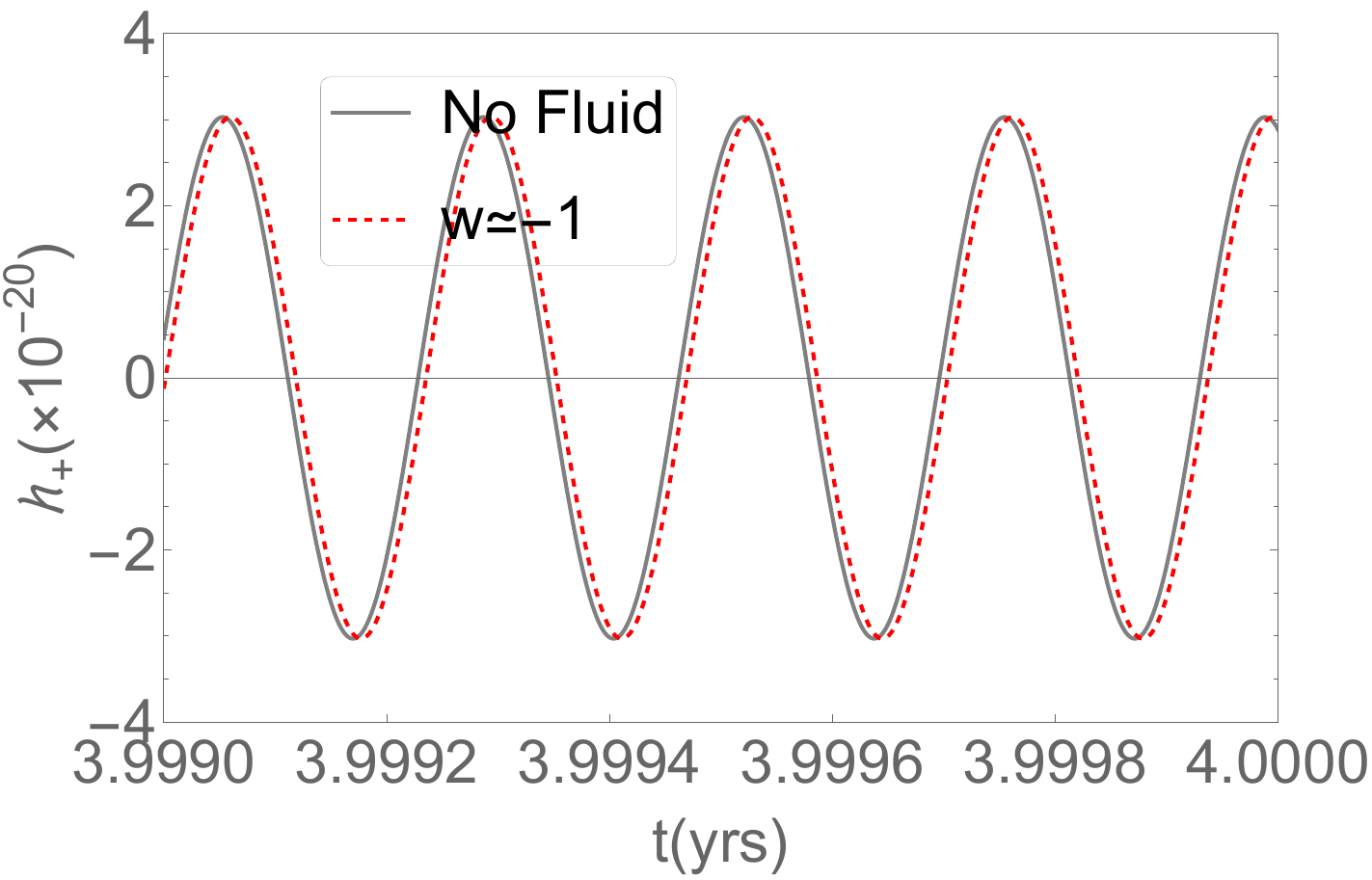}
\caption{ Upper Panel: Here, we present the de-phasing at 10 years induced by various accreting fluids. For this setup, we assume a background density \( \rho_{\infty} = \rho_{\rm crit} \equiv \frac{3 H_0^2}{8 \pi G} \). We consider a binary system initially located in a circular orbit with a radius \( r_{\text{init}} = 70 \, r_{\text{ISCO}} \) at \( t_{\text{init}} = 0 \). The orbital evolution is determined by solving Equations~(\ref{lagrangeeqr}) and (\ref{lagrangeeqphi}), incorporating the effects of the fluid through Equation~(\ref{frfluid}) and including the dynamical friction terms \( Q_r = F^r \) and \( Q_{\phi} = r F^{\phi} \) (see Equations~(\ref{dpdtr}) and (\ref{dpdtphi})).
This demonstrates how \(\Delta \phi\equiv \phi_{\rm fluid}-\phi\) (rad) at $10\,\text{yrs}$ varies as a function of the EoS parameter \(w\) at \(70r_{\rm ISCO}\). Lower Panel: We present the de-phasing for a binary system embedded in an accreting fluid, assuming a fluid density of \( \rho_{\infty} = 10^6 \rho_{\rm crit} \) with an equation of state parameter \( w \simeq -1 \), starting from an initial radius of \( r_{\rm init} = 10 \, r_{\rm ISCO} \). This is contrasted with the case where no surrounding fluid is present.  }
\label{fig:5a}
\end{figure}

    In the upper panel of Figure~\ref{fig:5a}, we present an example illustrating de-phasing over a 10-year timespan caused by different accreting fluids.  For this setup, we assume a background density \( \rho_{\infty} = \rho_{\rm crit} \equiv \frac{3 H_0^2}{8 \pi G} \). The binary system is positioned at \( \{R, \bar{\theta}, \bar{\phi}\} = \{1 \, \text{Mpc}, 0, 0\} \) with respect to the observer. The orbits are obtained by solving Equations~(\ref{lagrangeeqr}) and (\ref{lagrangeeqphi}) for several accreting fluids and substituting further into Equations~(\ref{quadropole}) and (\ref{hplus}) and (\ref{hx}). This illustrates how \( \Delta \phi \equiv \phi_{\text{fluid}} - \phi \) depends on the equation of state (EoS) parameter \( w \) at $70\, r_{\rm ISCO}$. In the lower panel of Figure~\ref{fig:5a}, we present an example, to roughly estimate the energy density required for the de-phasing to be potentially observable by future gravitational wave detectors, caused by global gravitational interaction. The de-phasing is shown for a binary system embedded in an accreting fluid with an equation of state parameter $w \simeq -1$, starting from an initial radius of $r_{\rm init} = 10 r_{\rm ISCO}$. This is contrasted with the case where no surrounding fluid is present. After 4 years, the de-phasing is $\Delta \phi \simeq -0.01\,\mathrm{rad}$, assuming a fluid density of $\rho_{\infty} = 10^6 \rho_{\rm crit}$.

    For instance, consider additional numerical examples: when the binary system is initialized at a distance of \( 250 \, r_{\rm ISCO} \), the de-phasing effects caused by a fluid with \( \rho_{\infty} = \rho_{\rm crit} \) and \( w \simeq -1 \) result in a phase difference of approximately \( \Delta \phi \equiv \phi_{\rm fluid} - \phi \simeq -3 \times 10^{-5} \, \text{rad} \) at the 10-year mark.
 Furthermore, to determine the required order of magnitude for the energy density necessary for dynamical friction to induce potentially observable de-phasing at such distances, consider a binary system with \( r_{\rm init} = 10 \, r_{\rm ISCO} \) embedded in an accreting fluid with \( \rho_{\infty} = 10^{25} \rho_{\rm crit} \) and equation of state parameters \( \alpha = 0.01 \) and \( \rho_0 = \rho_{\infty} \), corresponding to \( w(x=60) \simeq 0.007 \). In this case, the de-phasing at the 1-year mark, caused solely by dynamical friction while neglecting global gravitational interactions, is approximately \( \Delta \phi \simeq 0.001 \, \text{rad} \).

 As the separation between the binary components increases, the global radial force described by Equation~(\ref{frfluid}) becomes increasingly significant compared to dynamical friction, which depends only on local interactions with the fluid. 
 This is because a greater amount of energy density is enclosed within the spherical shell defined by the perturber's radius. Consequently, accretion has a negligible impact at such relative distances, and for the studied examples (Figure~\ref{fig:5a}) with parameter values of \(\alpha\) of order \(\sim\)0.1 (where \(\alpha\) determines the radial velocity \(u^r\) of the fluid at the critical point) its influence through dynamical friction becomes insignificant. For the spherical accretion of a fluid with \(\alpha\) parameter values in the order of \(\sim\)0.1 the dynamical friction on the perturber is negligible.
 However, in the current analysis, where \(m_2 \ll m_1\), non-negligible values at distances very close to the black hole horizon cannot be ruled out, as the PN approximation is not suitable for such situations \citep{Maggiore:2018sht}.
Thus, for spherically symmetric fluid accretion at relative distances far from the central supermassive black hole the dependence of the de-phasing on the EoS parameter \( w \) arises primarily through global gravitational interaction.

 A distinct phase shift, \( \Delta \phi \), is generally observed, varying with the fluid's equation of state  parameter \( w \). Notably, at the 10-year mark for a specific distance of \( 70 \, r_{\rm ISCO} \) the perturber in a binary without an accreting fluid exhibits a delay (\( \Delta \phi > 0 \)) relative to the corresponding perturber in a binary with an accreting fluid for \( w \gtrsim 0 \). Conversely, for \( w \lesssim 0 \) the perturber in a binary with an accreting fluid is delayed (\( \Delta \phi < 0 \)) compared to those without accreting fluid. This behavior depends on the energy density profile and the corresponding equation of state (EoS) that is integrated into Equation~(\ref{frfluid}).
 The specific value of \( w \) at which the transition from negative to positive \( \Delta \phi \) occurs varies with distance. 

The radial force \( f_{r}^{\rm fluid} \), as defined in Equation~(\ref{frfluid}), is considered a global force because it is computed as an integral over the region extending from the Schwarzschild radius to the radius of the perturber. For the specific distance of \( 70 r_{\rm ISCO} \) in the example considered, the absolute value of this force approaches zero for values of the EoS parameter \( w \) near \( 0 \) (upper panel of Figure~\ref{fig:5a}).
 However, the precise value of \(w\) at which \(\Delta \phi\) transitions from negative to positive depends on the distance, and it is identified by the point at which \(f_{r}^{\text{fluid}}\) becomes zero.

Although the effects on gravitational waveforms produced by an EMRI under the assumed low energy densities are negligible, they are expected to become significant in realistic astrophysical scenarios where higher densities are likely to be achieved. Despite being a preliminary study, this work demonstrates that the equation of state of the accreting energy density can be inferred from its impact on the binary system's dynamics.

\subsection{Effects of a Sudden Singularity on a Perturber via  Global Gravitational Interaction with \linebreak the Fluid}

In the present framework, the effects of cosmological singularities and their potential observable consequences for future gravitational wave telescopes can be effectively studied. In the context of General Relativity, various types of cosmological singularities exist \citep{deHaro:2023lbq}. A \textit{sudden cosmological singularity} or \textit{type II} singularity \citep{Barrow:2004xh} is geodesically complete and exhibits at the moment $a = a_{\rm s}<\infty$ a divergence of pressure $\big|p(a_{\rm s})\big| \to \infty$ with a finite density $\rho(a_{\rm s}) < \infty$, while the scale factor is continuous at $a_{\rm s}$. The sudden singularity can occur without violating the strong energy condition, even though the dominant energy condition is violated \citep{Barrow:2004xh}.

\subsection{Modeling the Impact of a Sudden Cosmological Singularity on a Black Hole}
The scale factor of a past sudden singularity event, ensuring that $a(\tau_0) = 1$, is parametrized as follows \citep{Paraskevas:2023aae}:

\begin{equation}\label{scalefactor}
   a(\tau) = \left(\frac{\tau_{\rm s}}{\tau_0}\right)^{\beta (1+\eta)} \left(\frac{\tau}{\tau_{\rm s}}\right)^{\beta[1+\eta \Theta(\tau-\tau_{\rm s})]}
\end{equation}
where $\eta$ is a dimensionless parameter and $\beta \equiv \frac{2}{3(1+w_{\infty})}$, with $w_{\infty}$ denoting the EoS parameter at infinity before the singularity event, far away from the central black hole. We assume the presence of non-phantom dark energy with \(-1 < w_{\infty} < -1/3\) ($1<\beta<\infty$), under which Equation~(\ref{scalefactor}) remains valid.
 The function $\Theta(\tau)$ represents the Heaviside step function \citep{ARFKEN20131}.
  Note that the second derivative of the scale factor diverges at the singularity. We apply the Friedmann equations:

\begin{equation}\label{rhoinfwiths}
   \rho_{\infty}(\tau)= \frac{3}{8 \pi G}H^2=\frac{3}{8\pi G}\left(\frac{\beta}{\tau}\right)^2\left[1+\eta\Theta(\tau-\tau_{\rm s})\right]^2,
\end{equation}
\begin{equation}\label{pinf}
    p_{\infty}(\tau)=-\frac{c^2}{8\pi G}\left(2\frac{\ddot{a}}{a}+H^2\right)=-\frac{c^2}{8\pi G}  \frac{(\beta + \beta \eta \, \Theta(\tau - \tau_s))^2}{\tau^2} - 
\end{equation}

    \begin{equation*}
         -\frac{c^2}{4\pi G} \left[ \frac{\beta \eta \, \delta(\tau - \tau_s)}{\tau} - 
    \frac{\beta (1 + \eta \, \Theta(\tau - \tau_s)) (-1 + \beta + 
          \beta \eta \, \Theta(\tau - \tau_s))}{\tau^2} \right]
     \end{equation*}
We simplify
\begin{equation}\label{rhopluspinf}
\rho_{\infty}+p_{\infty}/c^2=\frac{\beta\left[1+\eta\Theta(\tau-\tau_{\rm s})\right]}{4\pi G \tau^2}-\frac{\beta \eta}{4 \pi G \tau}\delta(\tau-\tau_{\rm s})
\end{equation}
where $\delta(\tau)$ represents the one-dimensional Dirac delta function \citep{ARFKEN20131}.

  As the pressure becomes infinite at the time of the sudden singularity, Equation~(\ref{dem0}) suggests that the time derivative of the black hole's mass will also diverge. As a result, a discontinuity in the rate of mass change is expected at the singularity.

The general solution of the differential Equation \eqref{dem0}, given Equation \eqref{rhopluspinf},  can be expressed as
\begin{equation}\label{massofbh}
    m(\tau)=\frac{1}{\frac{A G\beta}{c^3\tau}\left[1+\eta \Theta(\tau-\tau_{\rm s})\right]+\rm C}\,,\quad \text{C}=m_{\rm init}^{-1}-\frac{AG\beta }{c^3\tau_{\rm init}},
\end{equation}
given the initial condition \( m(\tau_{\rm init}) = m_{\rm init} \) and choosing the constant \( \rm C \) to be positive.
 The mass increases over time, assuming the presence of non-phantom dark energy ($-1<w_{\infty}<-1/3$), as discussed in \cite{Babichev:2004yx}. However, at the moment of the singularity, the mass undergoes a sudden decrease if $\eta > 0$, while it increases if $\eta < 0$ (see Equation~(\ref{rhopluspinf})). After the singularity event, the mass of the black hole resumes its growth.
\subsection{Effects on Geometry}

 The effect of the sudden singularity on the black hole's mass, as described by Equation~(\ref{massofbh}), is negligible. This is due to the fact that the term \(\sim [G/(c^3 \tau)]^{-1}\) over the lifetime of the universe remains negligible in comparison to \(m_{\rm init}\).   Consequently, the intensity of the singularity is also negligible, even for larger absolute values of \(\eta\) within our parametrization and example. This results in the black hole's mass staying practically constant  . Under this approximation, we will again apply Equations~(\ref{lagrangeeqr}) and (\ref{lagrangeeqphi}) in the following example, without accounting for variations in mass.

\subsection{An Example of a Stable Dark Fluid Model}
We begin with an equation of state (EoS) specified by $\alpha = 1$, and we adopt the profile derived by \citet{Babichev:2004yx}, which was obtained by analytically solving the system of Equations~(\ref{integr1lin}) and (\ref{integr2lin}). In this case, from Equation~(\ref{Aconstant}) we find that $A = 4$, and from Equations~(\ref{criticalvalues}), we have $x_{\rm c} = 2$ and $u^r_{\rm c}/c = -1/2$. These lead to the following expressions for the density profile and radial velocity as functions of $x$ \citep{Babichev:2004yx}:

\begin{equation}\label{examplefluidrho}
    \rho(x) =\rho_{\infty}\left[ q + \left( 1 -q \right) \left( 1 + \frac{2}{x} \right) \left( 1 + \frac{4}{x^2} \right)\right]
\end{equation}
where \(q \equiv \frac{\rho_0}{2 \rho_{\infty}}\). The corresponding radial fluid velocity, obtained from Equation~(\ref{integr1lin}), is given by
\begin{equation}\label{examplefluidur}
    \left(\frac{u^{r}}{c}\right)^2=\frac{16}{x(2+x)(4+x^2)}
\end{equation}
Note that the $\rho(x)$ and $u^r(x)$ of Equations~(\ref{examplefluidrho}) and (\ref{examplefluidur}) satisfy Equation~(\ref{enrgmomentumf}), as expected.
 For simplicity, we will study the impact of the singularity on the perturber solely through global gravitational interactions with the fluid. However, as is evident from the relations in Equations~(\ref{dpdtr}) and (\ref{dpdtphi}) for dynamical friction, the perturber will also experience the effects of the singularity through local gravitational interactions. Note that from Equations (\ref{integr2lin}) and (\ref{Aconstant}) and (\ref{examplefluidur}) we obtain
\begin{equation}\label{rhopluspresuuresing}
    \rho+p/c^2=\frac{(x+2)(4+x^2)}{x^3}\left(\rho_{\infty}+p_{\infty}/c^2\right)
\end{equation}

\begin{figure}[H]%
\includegraphics[width = 0.95\textwidth]{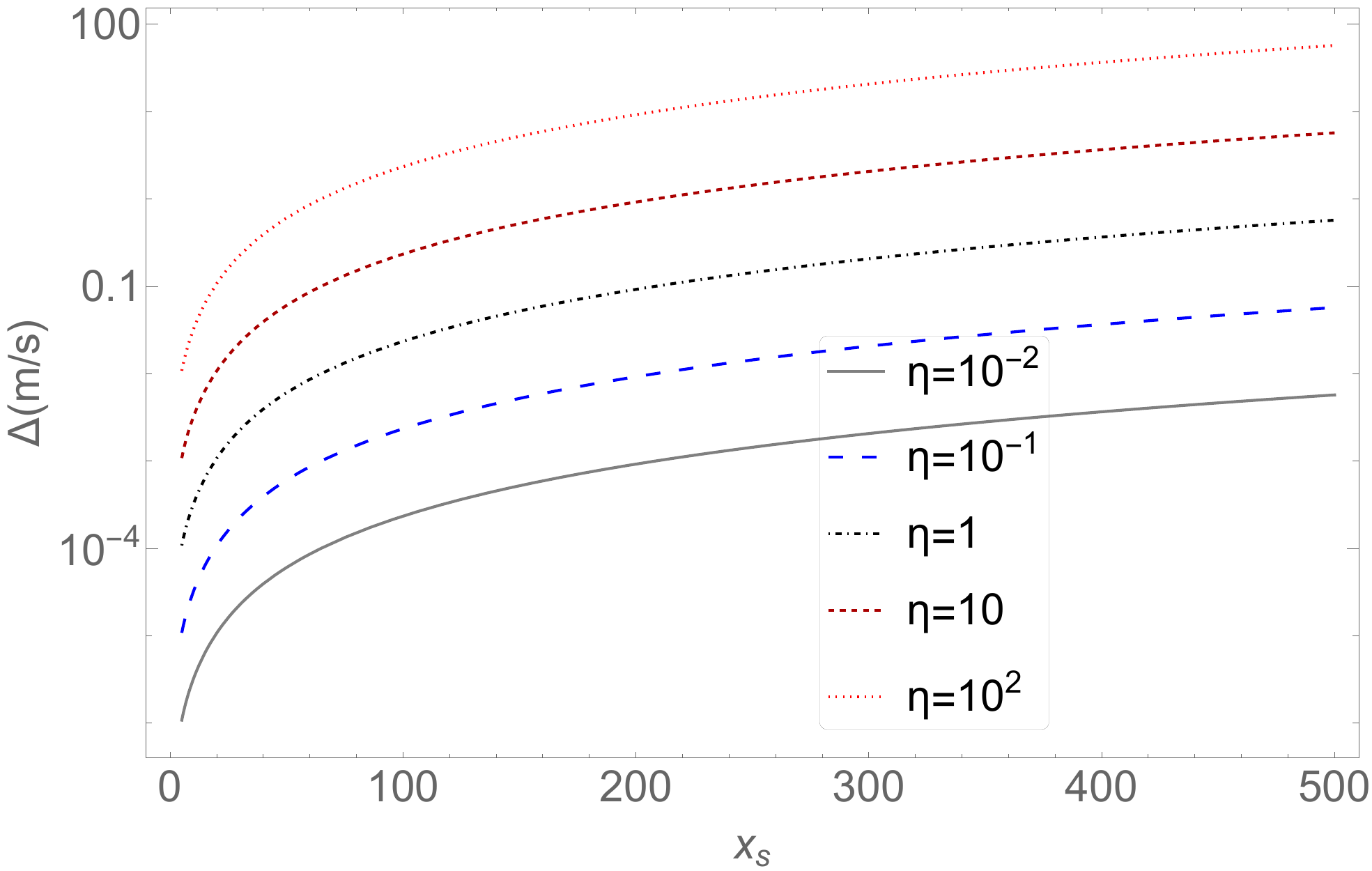}
\caption{The 
 impact of the sudden singularity induced by an accreting fluid with $\alpha = 1$ on the radial velocity of the perturber (see Equation~(\ref{velocitykick})) for $\tau_s =13.7 10^9\text{yrs}+  5 \, \text{yrs}$, $\beta = 666.667 \, (w_{\infty} = -0.999)$, and various parameter values of $\eta$, expressed as a function of the dimensionless parameter $x_s$ relative to the central black hole. }
\label{fig:kick}
\end{figure}

Consequently, we evaluate the gravitational interaction between the perturber and the fluid's density distribution inside a spherical region of radius \(r\), using the Poisson equation, similar to what was done in our earlier example in Equation~(\ref{frfluid}). This is achieved specifically by using Equations~(\ref{examplefluidrho}) and (\ref{rhopluspresuuresing}), given that we first substituted Equation~(\ref{rhoinfwiths}) into Equation~(\ref{examplefluidrho}) and Equation~(\ref{rhopluspinf}) into Equation~(\ref{rhopluspresuuresing}); then, we obtain for the radial gravitational interaction between the perturber and the fluid's density distribution inside a spherical region of radius \(r\),
\begin{equation}\label{frsing}
\begin{aligned}
    f_r^{\rm fluid} = & -\frac{4\pi G^2m}{c^2x^2}  \int_{2}^{x} dx'\, x'^2\left\{3\left(\rho_{\infty}(\tau) + \frac{p_{\infty}(\tau)}{c^2}\right)\frac{[x'+2][4+x'^2]}{x'^3}\right. \\
    &\left. - 2\rho_{\infty}\left[q - (1-q)\left(1 + \frac{2}{x'}\right)\left(1 + \frac{4}{x'^2}\right)\right]\right\} ,
\end{aligned}
\end{equation}
where $x$ is defined in Equation~(\ref{xdef}).   By integrating the differential Equation (\ref{lagrangeeqr}) over the time interval \([\tau_s - \epsilon, \tau_s + \epsilon]\) and taking the limit \(\epsilon \to 0\) we obtain the following result at \(\tau_s\):
\begin{equation}\label{velocitykick}
    \Delta \equiv\dot{r}_+-\dot{r}_-=\frac{3 Gm\beta \eta}{c^2x_s\tau_s}\left[\frac{x_s^3}{3}+x_s^2+4x_s-\frac{44}{3}+8\ln\left(\frac{x_s}{2}\right)\right]
\end{equation}
 As is evident from Equation~(\ref{velocitykick}), the magnitude of the velocity kick depends on \(w_{\infty}\), the jump parameter \(\eta\), the moment \(\tau_s\), and the distance $x_s$ at the moment $\tau_s$ from the black hole. The kick becomes more intense as we move further away from the black hole (see Fig.\ref{fig:kick}). 

\subsection{An Example of a Binary Black Hole System Enduring a Sudden Singularity}
We consider a small time interval (in contrast to the universe's cosmic expansion) around the singularity event, simplifying the analysis by neglecting the time dependence of the energy density. The binary system is assumed to be embedded in a universe where $\rho_{\infty}=\rho_{\rm \rm crit}$. The effects of the singularity event are accounted for through the velocity kick, as described in Equation~(\ref{velocitykick}), which serves as a boundary condition and occurs at $\tau_{\rm s} = 13.7 \times 10^9 \, \text{yrs} + t_s$, where $t_s = 5 \, \text{yrs}$. The value of $\tau_{\rm s}$ determines the magnitude of the velocity kick.
Additionally, after the singularity event ($t>t_{s}$) adjustments are made to $\rho_{\infty}$, $\rho_{\infty} + p_{\infty}/c^2$, and $f_{r}^{\rm fluid}$, as dictated by Equations~(\ref{rhoinfwiths})--(\ref{rhopluspinf}):
\[
\rho_{\infty} \to \rho_{\infty} \left[1 + \eta (\eta + 2)\right],\,\,\rho_{\infty} + \frac{p_{\infty}}{c^2} \to (1 + \eta) \left(\rho_{\infty} + \frac{p_{\infty}}{c^2}\right)
\]
\[
f^{\rm fluid}_{r}\to f^{\rm fluid}_{r} -4\pi Gr^{-2} \int_{r_S}^{r}\left\{ 3\eta\left[\rho(r) + \frac{p(r)}{c^2}\right]-2\eta(\eta+2)\rho(r)\right\}r^2 \, dr.
\]
Note that Equation~(\ref{integr2lin}) remains unaffected by the adjustment, as $\rho + p/c^2 \sim \rho_{\infty} + p_{\infty}/c^2$. Finally, we solve Equations~(\ref{lagrangeeqr})--(\ref{lagrangeeqphi}) in two separate regions: prior to and following the singularity event.

\begin{figure}[H]%

\includegraphics[width =0.9\textwidth]{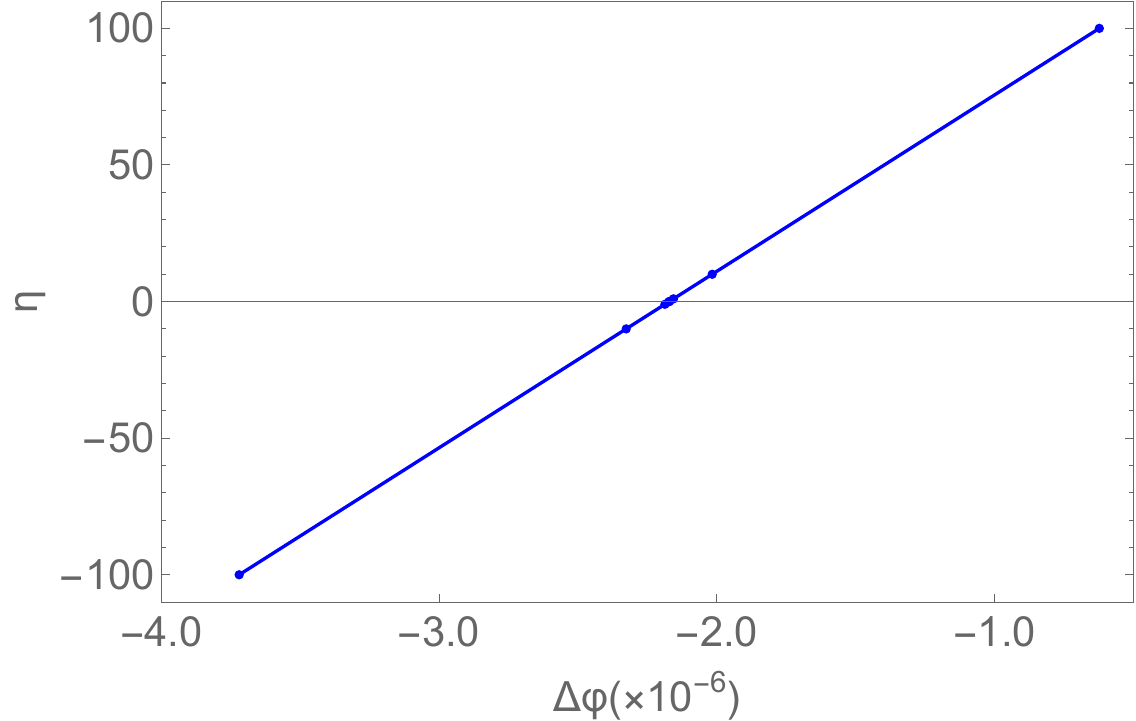}
\includegraphics[width = 0.9\textwidth]{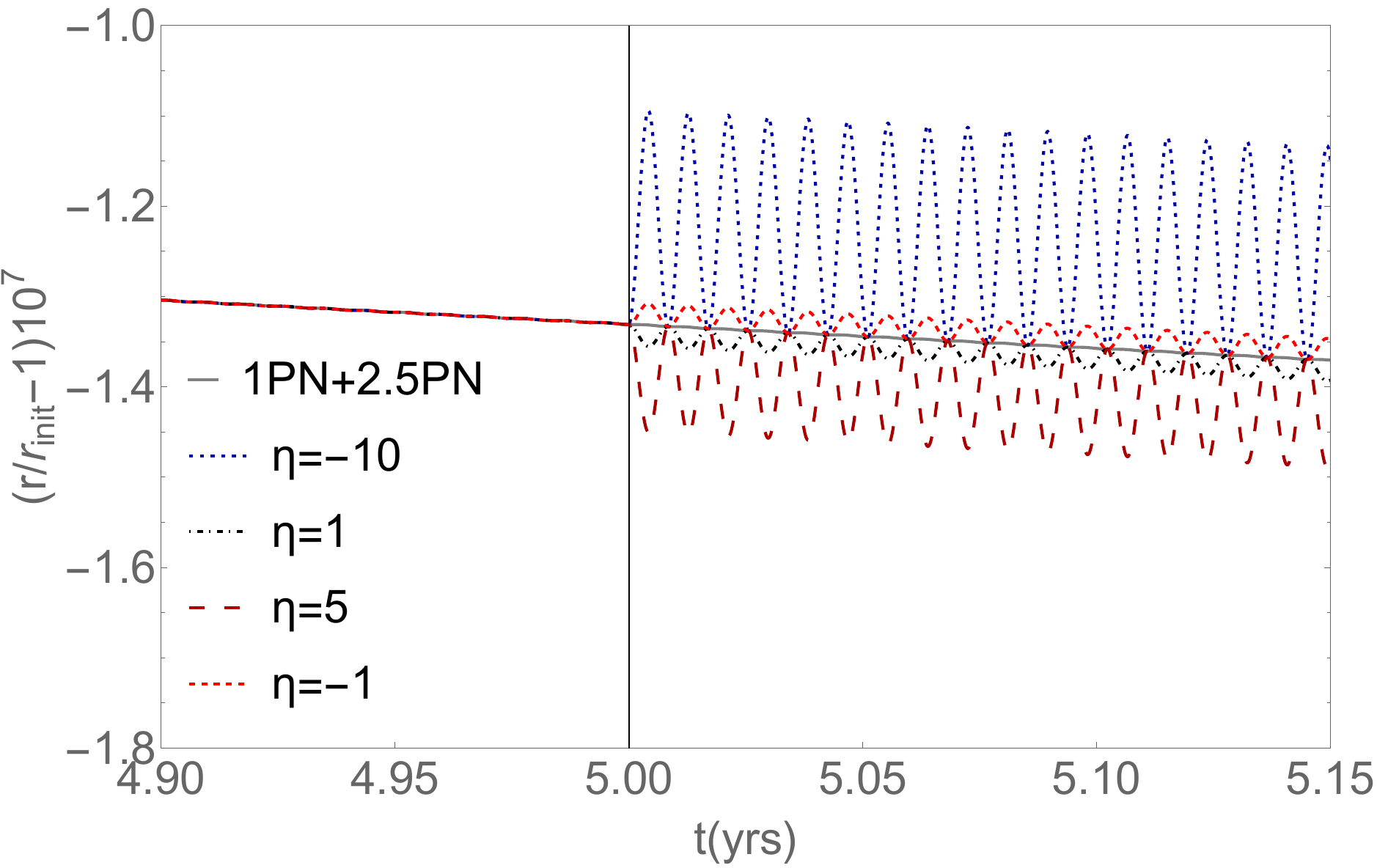}
\caption{We study a binary system with an initial circular orbit radius of $r_{\text{init}} = 70 \, r_{\text{ISCO}}$.   Post-Newtonian (PN) corrections at 1~PN and 2.5~PN are applied to the binary's orbit, which evolves within an accreting fluid characterized by an energy density $\rho_\infty = \rho_{\rm crit} = \frac{3H_0^2}{8\pi G}$. The fluid follows the equation of state $p = (\rho - \rho_0)c^2$, with $w_\infty = -0.999$.
Left Panel: The de-phasing $\Delta \phi \equiv \phi_{\rm fluid} - \phi$ at $5.01 \, \mathrm{yrs}$, induced by various sudden singularity events, is characterized by $\beta \simeq 666.667$ and different values of $\eta$ .
Right Panel: The percentage difference of the radius in terms of $r_{\rm init}$ during the time interval $[4.9 \, \text{yrs}, 5.2 \, \text{yrs}]$, expressed as $\left( \frac{r}{r_{\rm init}} - 1 \right) \times 10^7$, is presented for singularity events occurring at $t_s = 5 \, \text{yrs}$ for various values of $\eta = -10, -1, 1, 5$. The effect of these events on the radius $r(t)$ is demonstrated by comparing it to the scenario where no fluid is present. In all cases, we incorporate post-Newtonian (PN) corrections at 1~PN and 2.5~PN orders.}
\label{fig:8}
\end{figure}

In Figure~\ref{fig:8}, we study a binary system with an initial circular orbit radius of $r_{\text{init}} = 70 \, r_{\text{ISCO}}$.   Post-Newtonian (PN) corrections at 1~PN and 2.5~PN are applied to the binary's orbit, which evolves within an accreting fluid characterized by an energy density $\rho_\infty = \rho_{\rm crit} = \frac{3H_0^2}{8\pi G}$. The fluid follows the equation of state $p = (\rho - \rho_0)c^2$, with $w_\infty = -0.999$.
In the upper panel, we demonstrate  the de-phasing $\Delta \phi \equiv \phi_{\rm fluid} - \phi$ at $5.01 \, \mathrm{yrs}$, induced by various sudden singularity events, characterized by $\beta \simeq 666.667$ and different values of $\eta$.
In the lower panel, we show the percentage difference of the radius in terms of $r_{\rm init}$ during the time interval $[4.9 \, \text{yrs}, 5.2 \, \text{yrs}]$, expressed as $\left( \frac{r}{r_{\rm init}} - 1 \right) \times 10^7$, presented for singularity events occurring at $t_s = 5 \, \text{yrs}$ for various values of $\eta = -10, -1, 1, 5$. The effects of these events on the radius $r(t)$ are compared with the case where no singularity event occurs. In all cases, we incorporate post-Newtonian (PN) corrections at 1~PN and 2.5~PN orders.

\section{Conclusions}
In this study, we examined an extreme mass ratio inspiral (EMRI) binary system consisting of a supermassive black hole and a stellar-mass black hole. We analyzed the impact of an accreting dark fluid with a generic equation of state onto the supermassive black hole on the gravitational waveforms emitted by the binary. As long as \( m_1 \ll m_2 \), such that the center of mass approximately coincides with the center of the supermassive black hole and \( \mu \simeq m_2 \), the supermassive black hole can be modeled as a Schwarzschild black hole of mass \( m \), surrounded by a dark fluid at rest at infinity. The  steady-state flow of this fluid was studied, to determine its energy density and velocity profiles.
 The stellar-mass black hole acts as a perturber, interacting with the fluid through both local and global gravitational effects.

Subsequently, we examined the influence of the fluid on the binary system and its gravitational waveforms.
We began by validating the de-phasing results presented in \citet{Montalvo:2024iwq}, incorporating both 1~PN and 2.5~PN corrections, and we adopted the framework of \citet{Babichev:2013vji} to model steady-state flows of spherically symmetric accretion of fluids, including those with exotic equations of state resembling dark energy.  Our findings indicate that, at large separations, phase shifts in gravitational waveforms caused by spherically symmetric accretion of fluids are primarily driven by the global gravitational effects of the fluid on the perturber.

Additionally, we extended the analysis to examine the effects of sudden cosmological singularities \citep{Barrow:2004xh}. At the moment of the singularity, the pressure diverges to infinity, causing the time derivative of the black hole's mass to also diverge. This results in a discontinuity in the mass change. However, the intensity of this effect, even for higher values of $\eta$ in our parametrization and example, remains negligible, effectively keeping the black hole's mass practically constant.
While the change in mass is neglected in the equations of motion, the singularity can still influence the perturber's orbit through both global and local gravitational interactions with the fluid.
 For simplicity, and given its dominance at these relative distances, we focused on the global gravitational interaction. For cosmologically relevant energy densities, the velocity kick induced by the global gravitational interaction is negligible. Although the initial circular orbit may deform slightly into an elliptical one, the resulting eccentricity is insignificant, and the gravitational wave de-phasing caused by the singularity event is not potentially detectable.


Our results show a distinct relative de-phasing, \( \phi_{\rm fluid} - \phi \), induced by dark fluids compared to the case without a fluid. This de-phasing strongly depends on the fluid’s equation of state (EoS) through both local and global gravitational interactions with the perturber.  Notably, the de-phasing induced by the studied energy density and velocity profiles is negative for the EoS parameter \( w \lesssim 0 \) and positive for \( w \gtrsim 0 \) at a distance around 70 $r_{\rm ISCO}$.
This behavior may change, as the specific value of \( w \) at which the transition from negative to positive \( \Delta \phi \) occurs varies with distance.

For low energy densities, such as cosmologically relevant values (\( \rho_{\rm crit} \)), these effects may correlate with or be overshadowed by other factors, including post-Newtonian corrections, spin interactions, uncertainties in binary parameters, or deviations in the gravitational constant. However, at higher energy densities, these effects become significant. At high energy densities and in accretion scenarios resembling more realistic astrophysical conditions, such as non-spherically symmetric energy density profiles, dynamical friction may dominate the global gravitational interaction, potentially altering the relationship between \(w\) and \(\Delta \phi\).

These preliminary results demonstrate that the equation of state of the accreting fluid impacts gravitational waveforms. This work establishes a framework for understanding how fluid properties might be identified as observable features in gravitational waves from binary systems. Future studies should aim to improve the model by incorporating higher-order post-Newtonian terms, spin effects, and by studying the late inspiral and plunge phases.
However, numerical challenges associated with cases where $m_2 \ll m_1$ may complicate such efforts. Alternative methodologies should also be considered, to address these complexities. With the improved sensitivity of next-generation gravitational wave detectors, such research could reveal new insights into the nature of accreting fluids and their role in astrophysical and cosmological phenomena.

\vspace{6pt}

\authorcontributions{E.A.P. made contributions to the conceptualization, writing, methodology, and literature investigation. L.P. contributed to the conceptualization, writing, and general supervision of the project. All authors have thoroughly reviewed and approved the published version of \mbox{the manuscript}.}

\funding{This article is based upon work from COST Action CA21136---Addressing observational tensions
in cosmology with systematics and fundamental physics (CosmoVerse), supported by COST (European Cooperation in Science and Technology).
This project was also supported by the Hellenic Foundation for Research and Innovation (H.F.R.I.), under the ”First call for H.F.R.I. Research Projects to support Faculty members and Researchers and the
procurement of high-cost research equipment Grant”
(Project Number: 789).}

\dataavailability{This study is based on theoretical and symbolic calculations, with numerical computations performed using Mathematica where applicable. All relevant results and figures are included in the manuscript, and no external datasets were generated or used.}

\conflictsofinterest{The authors declare no conflicts of interest.}

\appendixtitles{yes}

\appendixstart

\appendix

\fontsize{12}{12}{\section[\appendixname~\thesection]{Perturber's Velocity in the Fluid Rest Frame Far from the Black Hole}\label{AppendixA}}

Assume a tetrad basis defined by the instantaneous rest frame of a particle in free fall in Schwarzschild spacetime, i.e., $(\mathbf{\hat{e}}_{a})^{\mu} (\mathbf{\hat{e}}_{b})^{\nu} g_{\mu\nu} = \eta_{ab}$, where the components are given \mbox{by \citep{Hobson:2006se}}
\begin{equation}
    (\mathbf{\hat{e}}_{\hat{t}})^{\mu}=\delta^{\mu}_t \left(1-\frac{2G m}{c^2r}\right)^{-\frac{1}{2}},\quad  (\mathbf{\hat{e}}_{\hat{r}})^{\mu}=\delta^{\mu}_r \left(1-\frac{2G m}{c^2r}\right)^{\frac{1}{2}},
\end{equation}
\begin{equation}
    (\mathbf{\hat{e}}_{\hat{\theta}})^{\mu}=\delta^{\mu}_\theta \frac{1}{r},\quad (\mathbf{\hat{e}}_{\hat{\phi}})^{\mu}=\delta^{\mu}_\phi \frac{1}{r \sin\theta}.
\end{equation}

Assume the perturber moves in a plane at $\theta=\frac{\pi}{2}$ along an almost circular arc, with a four-velocity given by $[v^\mu] = (v^t,v^r,v^\phi,0)$ in cylindrical coordinates. Additionally, assume that the worldlines of the perturber and an observer $\mathcal{O}$, who is falling with the fluid and has a four-velocity $u^\mu$, intersect at some event. The observer's four-velocity is given by $[u^\mu] = (u^t, u^r, 0, 0)$. At this event in spacetime, the velocities in the local frame will be given by $u^{\hat{b}} = u^{\mu} (\hat{e}^{b})_{\mu}$ and $v^{\hat{b}} = v^{\mu} (\hat{e}^{b})_{\mu}$. The $r$-component of the observer's ordinary velocity in the local tetrad frame is
\begin{equation}
    \frac{d\hat{r}}{d\hat{t}}\bigg|_{\mathcal{O}} = c\frac{u^{\hat{r}}}{u^{\hat{t}}} = c\frac{(\hat{e}^{\hat{r}})_{r}}{(\hat{e}^{\hat{t}})_{t}}\frac{u^{r}}{u^{t}} =c \left(1-\frac{2Gm}{c^2 r}\right)^{-1}\frac{u^{r}}{u^{t}}.
\end{equation}
Meanwhile, the four-velocity of the perturber in the local tetrad frame is
\begin{equation}\label{componentvelocitieslocalframe}
     v^{\hat{t}} = v^{t}\sqrt{1-\frac{2Gm}{c^2 r}},\quad v^{\hat{r}} = \frac{v^{r}}{\sqrt{1-\frac{2Gm}{c^2 r}}},\quad   v^{\hat{\phi}} =  v^\phi r,
\end{equation}
  The components of the perturber's ordinary velocity in the local tetrad reference frame are given by
 \begin{equation}\label{vrestframeinfinity}
    V^r = \frac{v^{\hat{r}}}{v^{\hat{t}}},\quad V^\phi = \frac{v^{\hat{\phi}}}{v^{\hat{t}}}.
\end{equation}

We approximate that at a sufficiently large distance from the central black hole the velocity components in the local frame, as expressed by Equation~(\ref{vrestframeinfinity}), will align with those of a stationary observer at infinity. Specifically, these components will approach $V^{r} \to \frac{dr}{dt}$ and $V^{\phi} \to \frac{d\phi}{dt} r$, meaning the velocity vector can be expressed as $\mathbf{V} = \dot{r} \mathbf{\hat{r}} + r \frac{d\phi}{dt} \boldsymbol{\hat{\phi}}$.
We assume that the observer $\mathcal{O}$ is located sufficiently far from the black hole, such that their coordinate $z$-axis aligns with that of a stationary observer at infinity. As $r \to \infty$, note that the observer $\mathcal{O}$ moves in the $\hat{r}$-direction with a velocity $\mathbf{w} = c\frac{u^r}{u^t}\hat{\mathbf{r}}$ (where the $u^t$ and $u^r$ components are calculated in section \ref{accretingfluid}), as observed by a stationary observer at infinity.

To obtain the components of the ordinary three-velocity of the perturber in the rest frame of the observer $\mathcal{O}$, we apply the relativistic transformation of velocities. Under this transformation, the components are given by \citep{Jackson:1998nia}
\begin{equation}\label{capitalv}
    \tilde{V}^r = \frac{\frac{dr}{dt}-w}{1-\frac{w\frac{dr}{dt}}{c^2}},\quad \tilde{V}^\phi = \frac{r\frac{d\phi}{dt}}{\gamma_{w}\left(1-\frac{w\frac{dr}{dt}}{c^2}\right)}.
\end{equation}
Thus, the velocity magnitude, as seen by an observer moving with the fluid, is given by $\tilde{V} = \sqrt{(\tilde{V}^r)^2 + (\tilde{V}^\phi)^2}$ at the point where the worldlines of the falling observer (with the fluid) and the perturber intersect.

\begin{adjustwidth}{-\extralength}{0cm}
\printendnotes[custom]
\reftitle{References}

\PublishersNote{}
\end{adjustwidth}

\end{document}